\documentclass[a4paper,fleqn,usenatbib]{mnras}

\usepackage{newtxtext,newtxmath}

\usepackage[T1]{fontenc}
\usepackage{ae,aecompl}
\usepackage{float}
\usepackage{multirow}

\usepackage{graphicx}	
\usepackage{amsmath}	
\usepackage{amssymb}	
\usepackage{natbib}

\title[Instabilities in primordial stars]{Strange mode instabilities and mass-loss in evolved massive primordial stars}

\author[A. P. Yadav, S. K\"uhnrich and W. Glatzel]
{Abhay Pratap Yadav,\thanks{E-mail:yadav@astro.physik.uni-goettingen.de}
Stefan Henrique K\"uhnrich Biavatti and Wolfgang Glatzel 
\\
Institut f\"ur Astrophysik (IAG), Georg-August-Universit\"at G\"ottingen, 
              Friedrich-Hund-Platz 1, D-37077 G\"ottingen, Germany
}

\date{Accepted XXX. Received YYY; in original form ZZZ}

\pubyear{2018}

\begin{document}
\label{firstpage}
\pagerange{\pageref{firstpage}--\pageref{lastpage}}
\maketitle

\begin{abstract}

 A linear stability analysis of models for evolved primordial stars with masses between 150 and 250 M$_{\sun}$ is presented. Strange mode 
 instabilities with growth rates in the dynamical range are identified for stellar models with effective temperatures below 
 log T$_{\rm{eff}}$ = 4.5. For selected models the final fate of the instabilities is determined by numerical simulation of their evolution into 
 the non-linear regime. As a result, the instabilities lead to finite amplitude pulsations. Associated with them are acoustic energy fluxes capable 
 of driving stellar winds with mass-loss rates in the range between 7.7 $\times$ 10$^{-7}$ and 3.5 $\times$ 10$^{-4}$ M$_{\sun}$ yr$^{-1}$.

\end{abstract}




\section{Introduction}

Primordial stars with initially vanishing metallicity 
are cosmologically relevant in many
respects: Being the only
source for the production of metals they are responsible 
for chemical evolution in the early universe \citep[see][]{nomoto_2013}. Thus they have an important
influence on the formation and evolution of cosmic structure.
Having masses up to $\sim$ 1000 M$_\odot$ they typically
end their life as pair instability supernovae enriching
their environment with heavy elements or leaving behind
intermediate mass black holes \citep[see, e.g.,][]{bromm_2004, ohkubo_2009}. Being smaller and hotter
than their counterparts with finite metallicity they are
the most promising source of the reionization of the universe \citep[see, e.g.,][]{tumlinson_2000, barkana_2001, loeb_2001}.
Indication for the existence of primordial (Pop III) stars 
has meanwhile been found from the observations of high redshift
galaxies \citep[see, e.g.,][]{fosbury_2003, kashlinsky_2005}. The study presented in this paper of the structure 
and evolution of primordial stars including consideration of their 
stability is therefore of
utmost importance for cosmology. Should an instability prevail
which leads to pulsationally driven mass loss, it would be
relevant not only for the evolution of Pop III stars but also 
for the enrichment with metals of the environment.

Investigations of the formation process of primordial stars indicate that the absence of metals allows for the formation of 
massive fragments with masses in the range between 100 and 1000 M$_{\sun}$ \citep[see, e.g.,][]{abel_2000, bromm_1999, bromm_2002}.
Whether these massive primordial Pop III stars suffer from 
significant mass-loss during their evolution or the evolution proceeds at constant mass, is still a matter of debate. Due to the absence 
of metals, line driven winds can be excluded as a source of mass-loss. 
A stability analysis of Pop III stars with respect to the $\varepsilon$ - 
mechanism revealed instabilities which are too weak to drive a significant mass-loss \citep{baraffe_2001}.   
Moreover, this $\varepsilon$ - instability is restricted to the very vicinity of the zero age main sequence.
On the other hand, massive Pop III stars are characterized by high luminosity to mass ratios (>10$^{3}$ in solar units) which imply 
a high fraction of the radiation pressure in the envelopes of these stars. Both high L/M ratios and dominant radiation pressure 
favour the occurrence of strange mode instabilities \citep[see, e.g.,][]{glatzel_1994}. 
Therefore strange mode instabilities are to be expected in massive primordial stars.

The objective of the present study is to identify strange mode instabilities in massive primordial stars by a linear stability analysis 
and subsequently to determine the final result by numerical simulation of their evolution into the non-linear regime. Evolution and stability
with respect to the $\varepsilon$ - mechanism close to the main sequence has been studied by \citet{baraffe_2001}. In the present study we shall 
therefore ignore this phase and restrict our investigation to the post main sequence phase, where the evolution proceeds at almost constant mass
and luminosity from high to low effective temperature. Moreover, since the $\varepsilon$ - mechanism is disregarded and the strange mode instabilities
of interest operate in the stellar envelope only, we can restrict ourselves to the consideration of envelopes.

The stellar models considered will be described in section 2, their linear stability analysis in sections 3 and 4. Non-linear simulation are discussed 
in section 5. A discussion and our conclusions follow (section 6).


   \section{Models}
   
   Concerning the objective to study strange mode instability of evolved massive primordial stars disregarding  $\varepsilon$ - instability, 
   we restrict ourselves to the investigation of envelope models (rotation and magnetic fields are ignored) with masses of
   150, 200 and 250 M$_{\sun}$, respectively. In the post main sequence phase, evolution proceeds at almost constant mass and luminosity. 
   Thus these masses correspond to a luminosity of log L/L$_{\sun}$ = 6.60, 6.77 and 6.88, respectively \citep[see][]{moriya_2015, baraffe_2001}. 
   The effective temperature is varied between log T$_{\rm{eff}}$ = 4.80 and log T$_{\rm{eff}}$ = 3.62. For the chemical composition, we adopt 
   primordial values Z = 0.00, X = 0.77 and Y = 0.23. Opacities are taken from the OPAL tables \citep{rogers_1992, rogers_1996, iglesias_1996}.

   For given mass, luminosity and effective temperature envelope models are constructed by integrating the equations of mass conservation, hydrostatic
   equilibrium and energy transport from the photosphere to a maximum cutoff temperature. To ensure that the parts of the envelope relevant
   for stability are represented, the latter has to be chosen sufficiently high. 
   For the initial conditions of the integration, we have adopted Stefan-Boltzmann's law and the common prescription for
   the photospheric pressure \citep[see section 11.2 of][]{kippenhahn_2012}.

\begin{figure}
\centering $
\Large
\begin{array}{c}
  \scalebox{0.68}{ \input{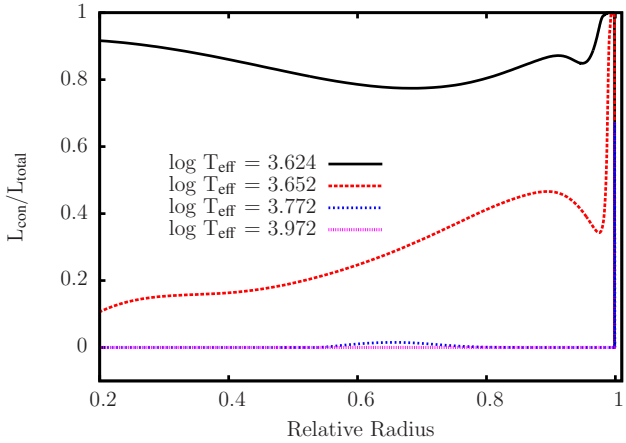} } \\
 \end{array}$
 \caption{ Ratio of convective and total luminosity  as a function of relative radius for stellar 
  models with M = 150 M$_{\sun}$ and log L/L$_{\sun}$ = 6.6 and the effective temperatures indicated.
  With decreasing effective temperature, the contribution of convection to the energy transport in the envelope increases significantly. }
  
 \normalsize
 \label{ratio}
 \end{figure}

   Concerning the energy transport, Schwarzschild's criterion has been used for the onset of convection. Convection is treated according to the 
   standard mixing length theory \citep{bohm_1958} with  1.6 pressure scale heights for the mixing length. In models with effective temperatures 
   above log T$_{\rm{eff}}$ = 3.7 energy transport by convection in negligible, below log T$_{\rm{eff}}$ = 3.7, the fraction of convectively 
   transported energy strongly increases with decreasing T$_{\rm{eff}}$. This is illustrated in Fig. \ref{ratio}, where the ratio of the 
   convective and the total luminosity is given as a function of relative radius for stellar models with different effective temperatures. 
   Stellar models with log T$_{\rm{eff}}$ $\approx$ 3.6 are fully convective. 
   


   \section{ Linear Stability analysis}
  
  The linear stability analysis is based on the equations governing linear stability and pulsations in the form given by 
   \citet[][equation 2.12]{gautschy_1990b}. Together with four boundary conditions, they form a fourth order
  eigenvalue problem. It is solved using the Riccati method introduced by \citet{gautschy_1990a}.
  The eigenvalues are complex where the real parts ($\sigma_{\rm{r}}$) correspond to
  the pulsation frequency and the imaginary 
  parts ($\sigma_{\rm{i}}$) provide information about excitation or damping of the corresponding mode.
  Negative values of the imaginary part ($\sigma_{\rm{i}}$ < 0) indicate excitation and instability, positive values ($\sigma_{\rm{i}}$ > 0) 
  correspond to damping. In the following, eigenvalues will be normalized by the global free fall time 
  $\sqrt{\frac{R^{3}}{3GM}}$ ( $G$, $R$ and $M$ are the gravitational constant, radius and mass of the stellar model considered, respectively). 
  For the normalization see also \citet{baker_1962}.

  As a theory of the interaction of pulsation and convection is still not available, we have adopted for the treatment of convection 
  the `frozen in approximation' introduced by \cite{baker_1965}. In this approximation, the Lagrangian perturbation of the convective luminosity
  is disregarded in the pulsation equations. It is applicable as long as the convective turn over timescale is longer than the pulsation timescale
  and if the energy is mainly transported by radiation diffusion \citep[see][for a detailed discussion]{baker_1965}.
  For the models considered here, the frozen in approximation holds for log T$_{\rm{eff}}$ > 3.7. However, as discussed in the previous section
  (see Fig. \ref{ratio}) below log T$_{\rm{eff}}$ $\approx$ 3.7 convection is dominant and the results of the stability analysis have to be 
  interpreted with caution.    
  


 \begin{figure*}
\centering $
\Large
\begin{array}{cc}
\scalebox{0.68}{ \input{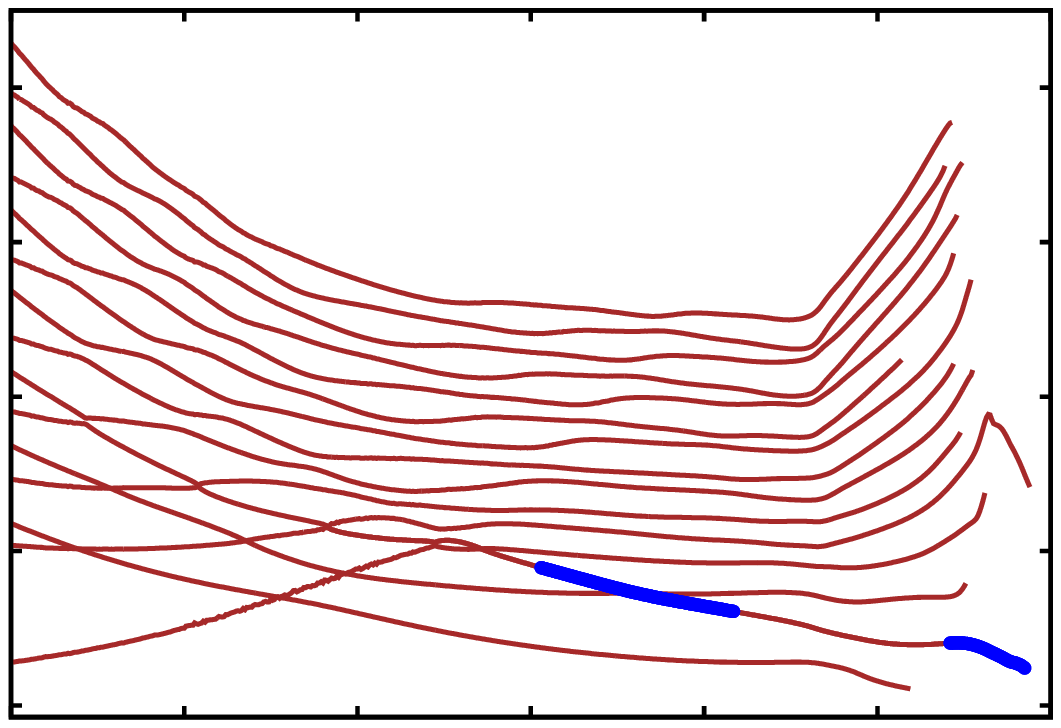} } 
   \scalebox{0.68}{ \input{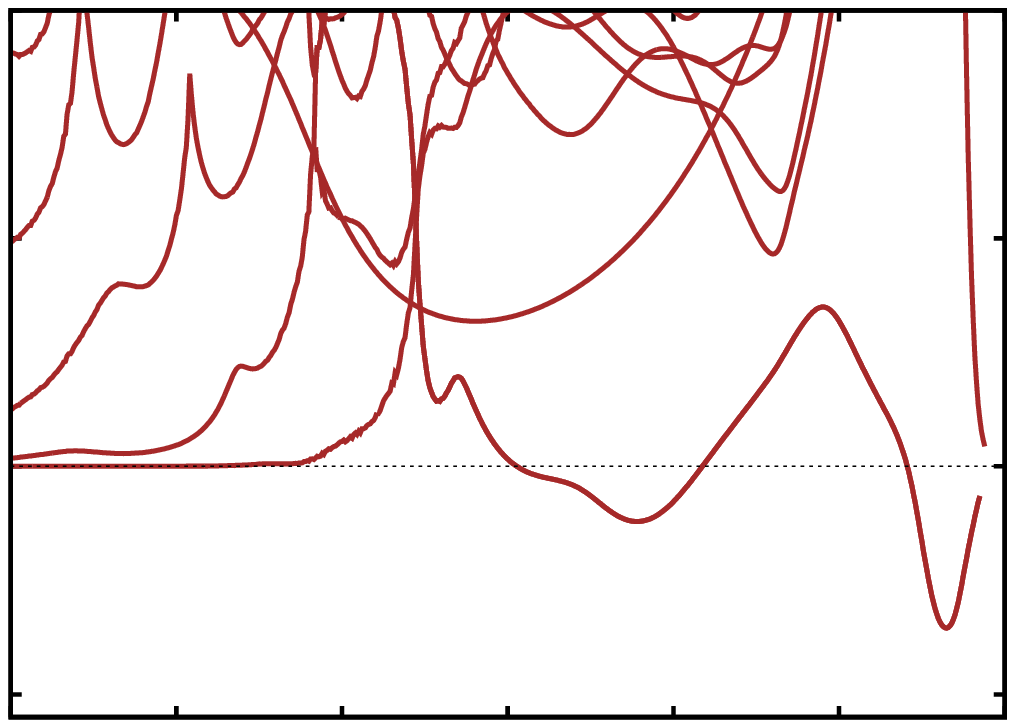} } \\
 \end{array}$
 \caption{Real (a) and imaginary parts (b) of the eigenfrequencies normalized by the 
 global free fall time as a function of effective temperature for stellar models with M = 150 M$_{\sun}$ and log L/L$_{\sun}$ = 6.6.
 Unstable modes are indicated by thick lines in (a)
 and negative imaginary parts in (b). For the photospheric boundary conditions vanishing Lagrangian pressure perturbation 
 and Stefan Boltzmann's law have been adopted.}
 \normalsize
 \label{150_modal_OBC}
 \end{figure*}

  \begin{figure*}
\centering $
\Large
\begin{array}{cc}
  \scalebox{0.68}{ \input{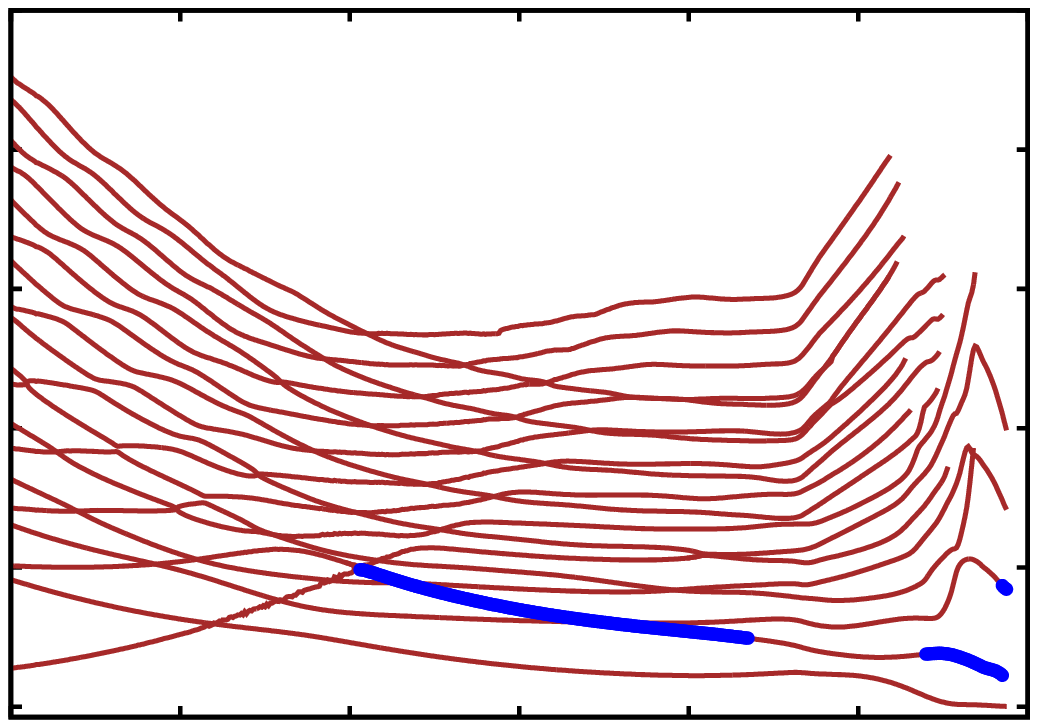} } 
   \scalebox{0.68}{ \input{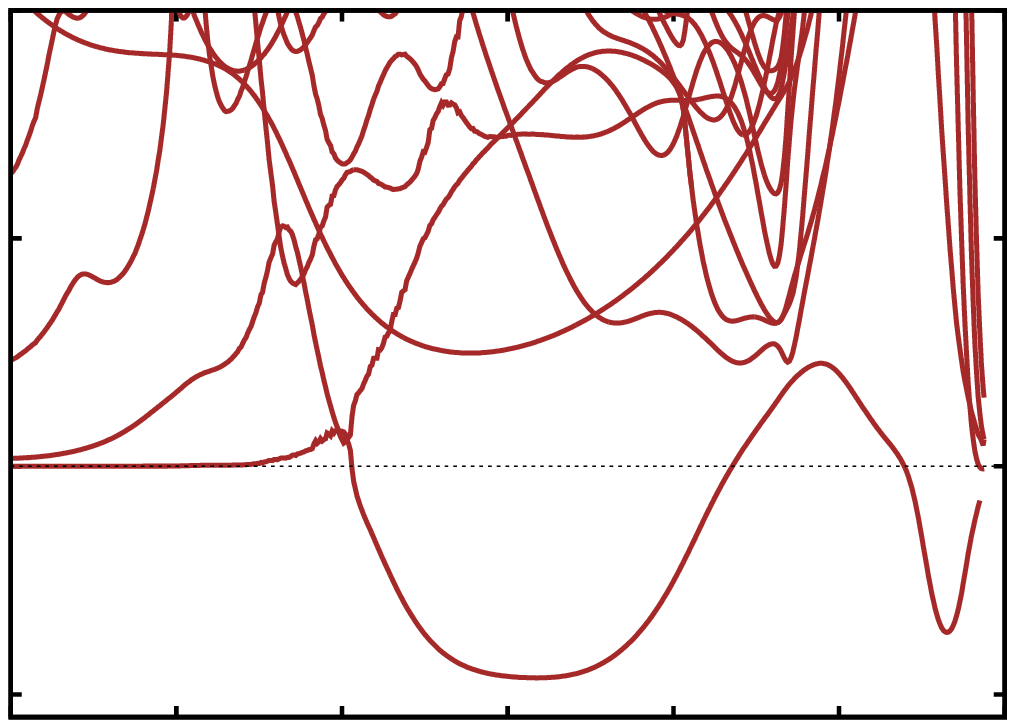} } \\
 \end{array}$
 \caption{ Same as Fig. \ref{150_modal_OBC} but for M = 200 M$_{\sun}$ and log L/L$_{\sun}$ = 6.77.}
 \normalsize
 \label{200_modal_OBC}
 \end{figure*}

   \begin{figure*}
\centering $
\Large
\begin{array}{cc}
   \scalebox{0.68}{ \input{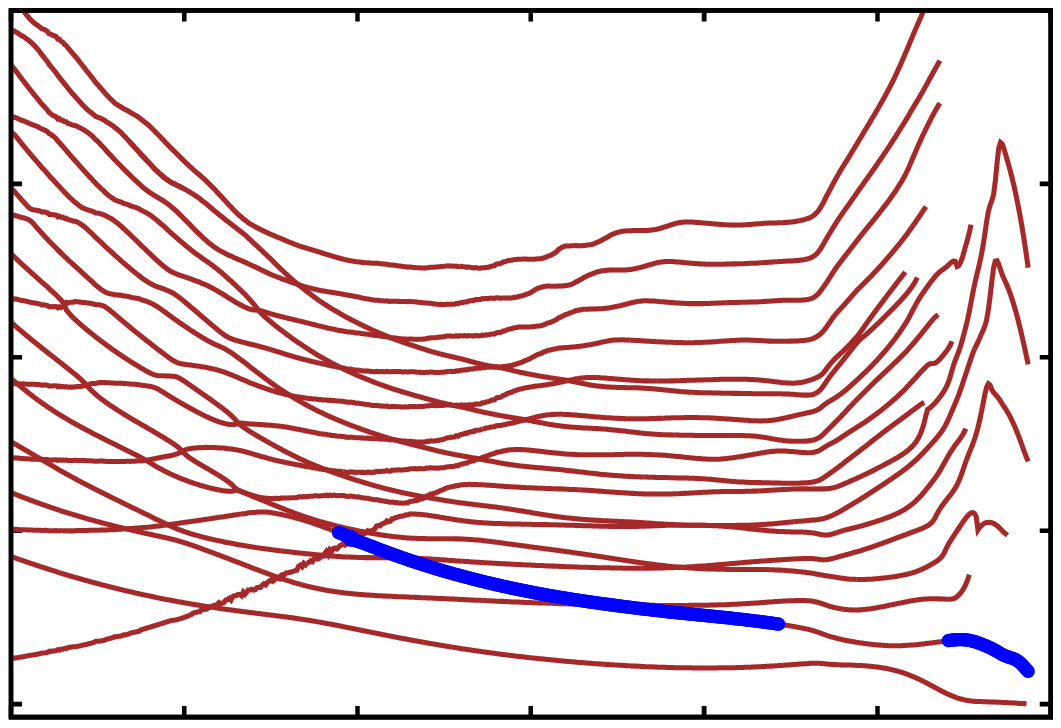} } 
    \scalebox{0.68}{ \input{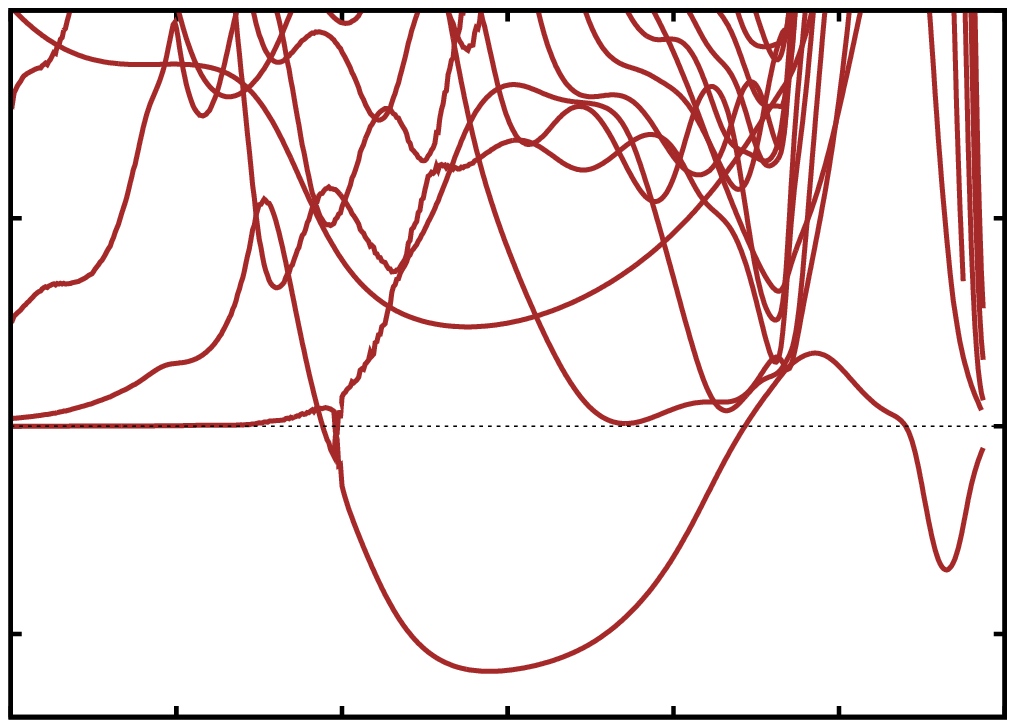} } \\
   
 \end{array}$
 \caption{ Same as Fig. \ref{150_modal_OBC} but for M = 250 M$_{\sun}$ and log L/L$_{\sun}$ = 6.88.}
 \normalsize
 \label{250_modal_OBC}
 \end{figure*}

\begin{figure*}
\centering $
\Large
\begin{array}{cc}
  \scalebox{0.68}{ \input{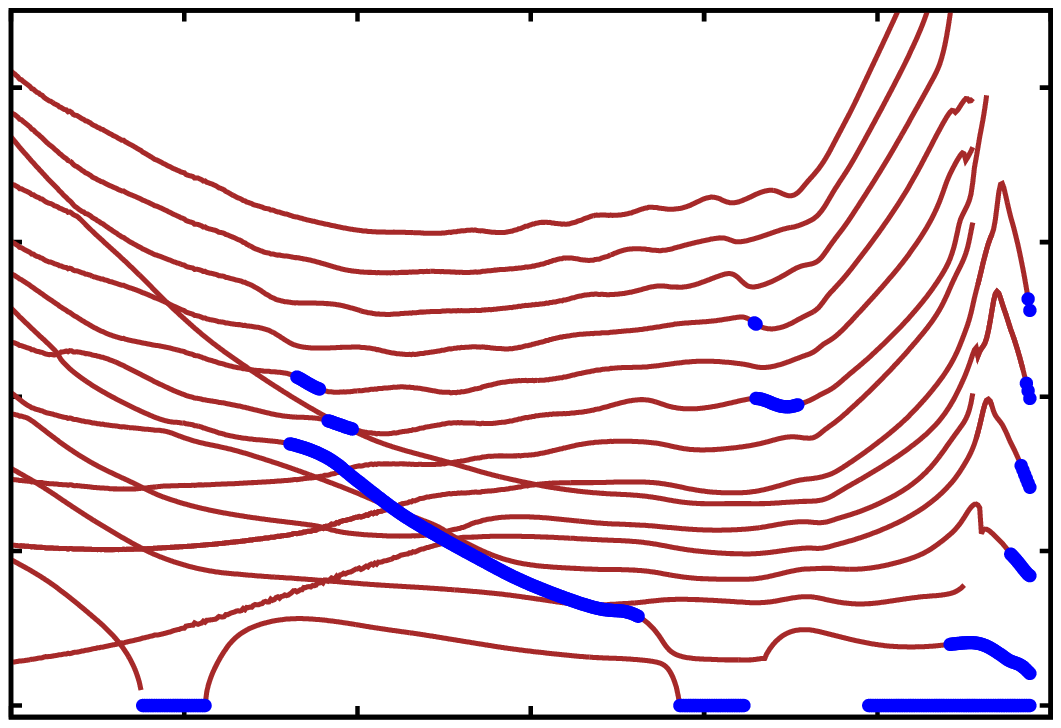} } 
   \scalebox{0.68}{ \input{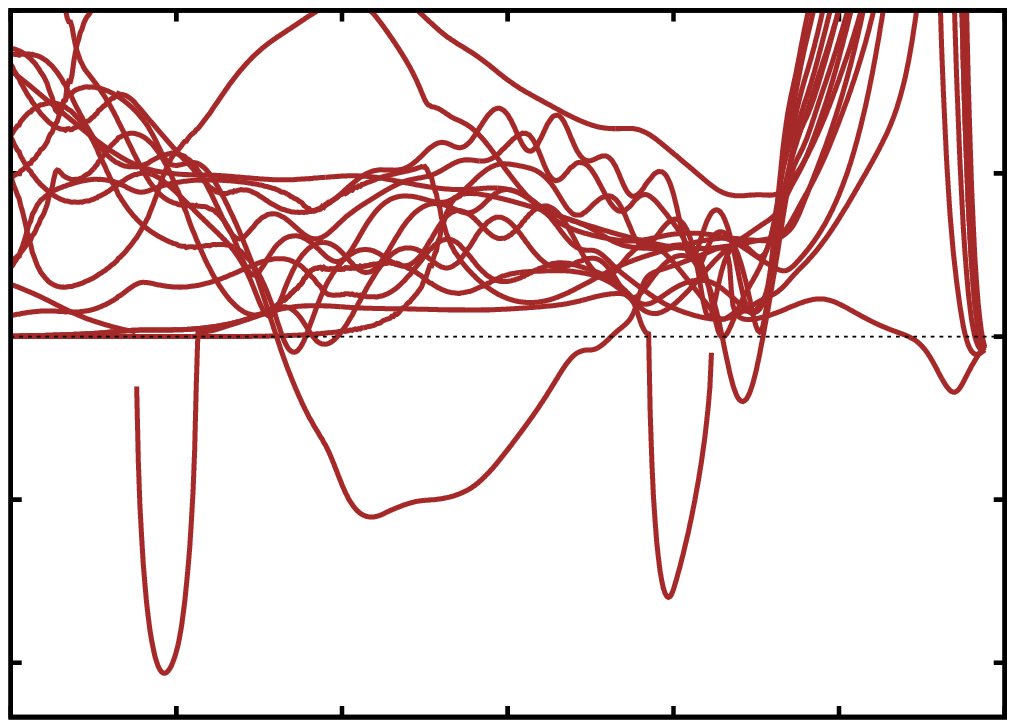} } \\  
 \end{array}$
 \caption{ Same as Fig. \ref{150_modal_OBC} but for photospheric boundary conditions requiring the gradient of compression and 
 the divergence of the heat flux to vanish.}
 \normalsize
 \label{150m_modal_nlb}
 \end{figure*}

  \begin{figure*}
\centering $
\Large
\begin{array}{cc}
   \scalebox{0.68}{ \input{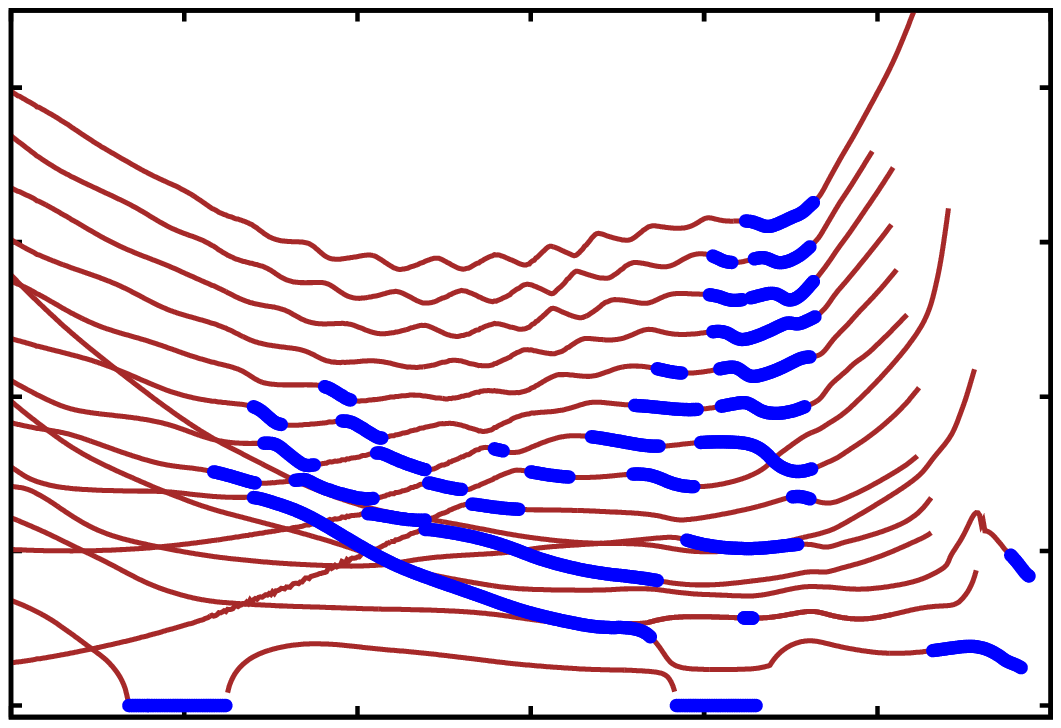} } 
    \scalebox{0.68}{ \input{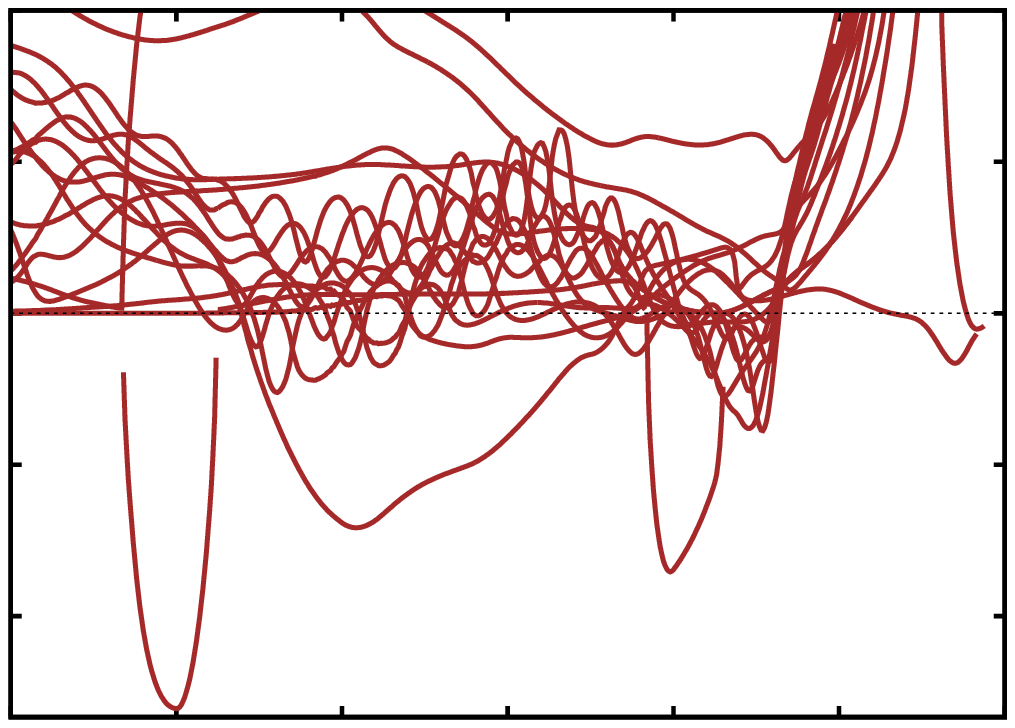} } 
 \end{array}$
 \caption{Same as Fig. \ref{150m_modal_nlb} but for M = 200 M$_{\sun}$ and log L/L$_{\sun}$ = 6.77.}
 \normalsize
 \label{200m_modal_nlb}
 \end{figure*}

  \begin{figure*}
\centering $
\Large
\begin{array}{cc}

   \scalebox{0.68}{ \input{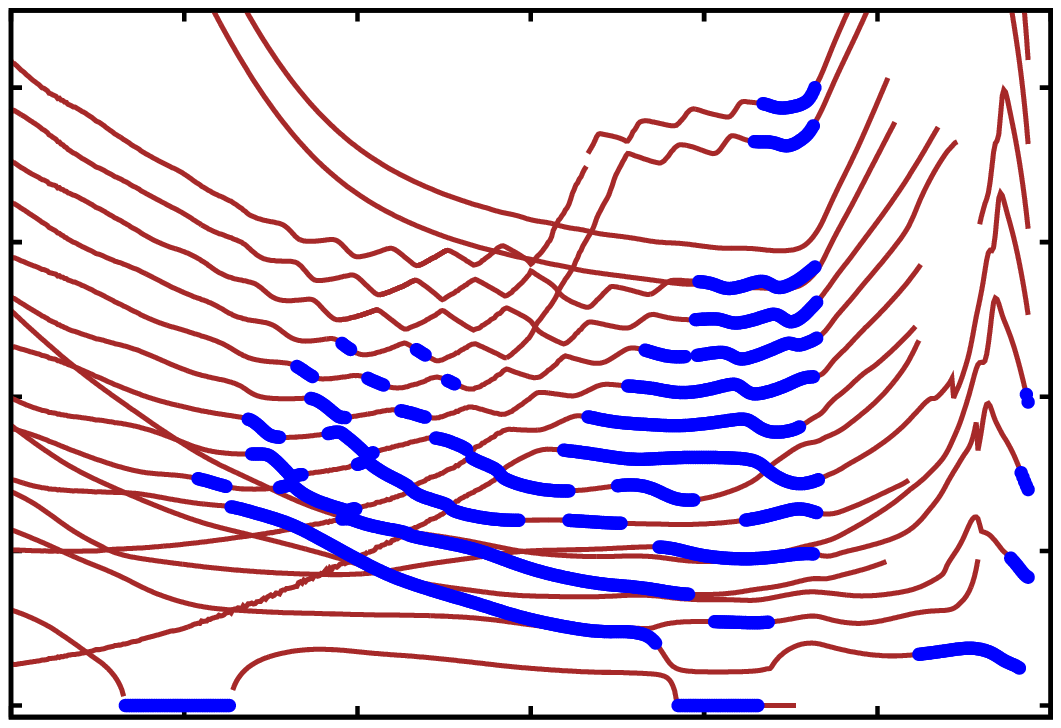} } 
    \scalebox{0.68}{ \input{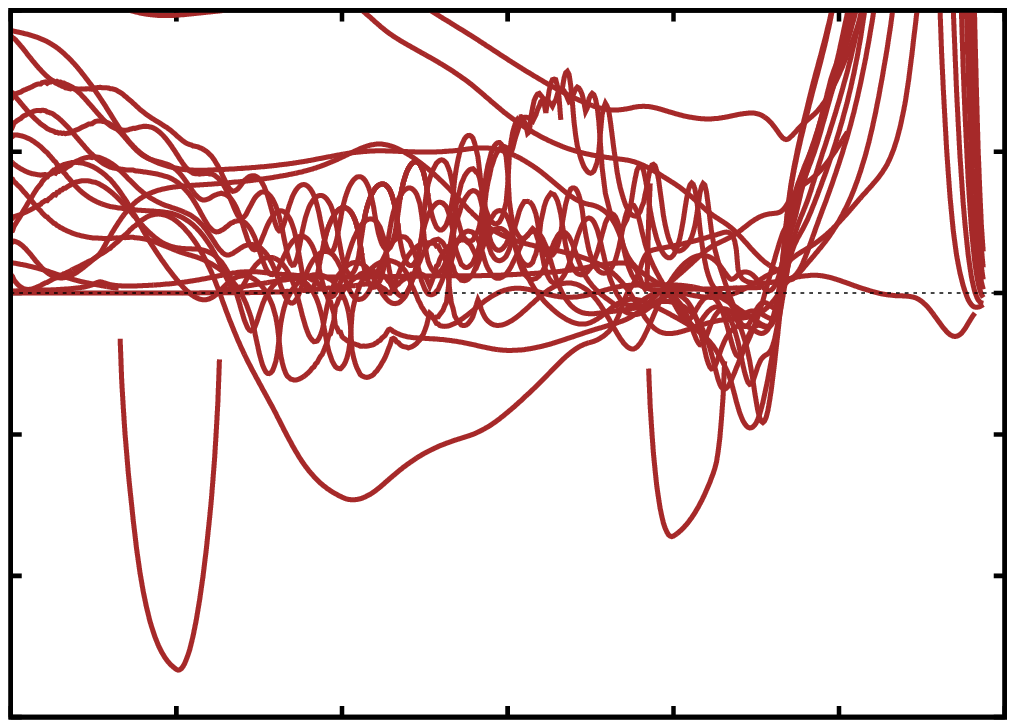} } 
 \end{array}$
 \caption{Same as Fig. \ref{150m_modal_nlb} but for M = 250 M$_{\sun}$ and log L/L$_{\sun}$ = 6.88.}
 \normalsize
 \label{250m_modal_nlb}
 \end{figure*}

\section{Results of the linear stability analysis} 
For the three masses  150, 200 and 250 M$_{\sun}$ and associated luminosities log L/L$_{\sun}$ = 6.60, 6.77 and 6.88, the results of the 
linear stability analysis, i.e., the real and imaginary parts ($\sigma_{r}$, $\sigma_{i}$) of the eigenvalues normalized by the global 
free fall time are presented as a function of the effective temperature in Figs. \ref{150_modal_OBC} - \ref{250m_modal_nlb} which will be 
referred to as modal diagrams in the following. As long as the bottom boundary of the envelope is chosen sufficiently deep, the 
eigenfrequencies are neither sensitive to its position nor to the boundary conditions imposed there, which is a consequence of the 
common exponential decay of the eigenfunctions from the surface to the stellar core. For the boundary conditions at the photosphere, we have 
considered the conventional set, where the Lagrangian pressure perturbation is required to vanish and a linearized version of 
Stefan - Boltzmann's law is assumed to hold \citep[see][]{baker_1962}. Alternatively, we have chosen boundary conditions which are
identical with those used in the subsequent non-linear simulations. These boundary conditions requiring the gradient of compression and 
the divergence of the heat flux to vanish are constructed such that reflection of waves and shocks at the outer boundary is minimized 
\citep[see][]{grott_2005}. For stellar models previously tested for linear stability, the choice of the photospheric boundary  
conditions was not crucial \citep{yadav_2017, yadav_2017b}. Even if occasionally quantitative differences were observed, the results of the 
linear stability analysis did qualitatively not depend on the outer boundary conditions. For the massive primordial models considered here 
we have performed a stability analysis both using the conventional outer boundary conditions (results are shown in the modal diagrams 
Figs. \ref{150_modal_OBC} - \ref{250_modal_OBC}) and with the boundary conditions consistent with the subsequent non-linear treatment 
 (results are shown in the modal diagrams Figs. \ref{150m_modal_nlb} - \ref{250m_modal_nlb}).

 The stability analysis of models with 150 M$_{\sun}$ using conventional boundary conditions (Fig. \ref{150_modal_OBC}) exhibits a 
 complex behaviour of the eigenvalues which is typical for the expected strange mode phenomenon 
 \citep[see, e.g.,][]{yadav_2016, yadav_2017, yadav_2017b}: At least two sets of modes can be identified by actual crossings and sequences 
 of avoided crossings. One of the mode crossings has unfolded into an instability band implying instabilities with growth 
 rates in the dynamical 
 regime for effective temperatures between log T$_{\rm{eff}}$ = 4.21 and log T$_{\rm{eff}}$ = 3.98. A second instability associated with 
 the second lowest eigenfrequency $\sigma_{r}$ is observed for effective temperatures below  log T$_{\rm{eff}}$ = 3.7 
 (see Fig. \ref{150_modal_OBC}). We emphasize that a classification in terms of fundamental modes and overtones is not applicable 
 here due to significant deviations from adiabatic behaviour. In models with log T$_{\rm{eff}}$ < 3.7, energy transport is dominated by  
 convection. Thus the results of the stability analysis in this range, in particular the instability identified there, have to be 
 interpreted with caution. Contrariwise, the strange mode instability for effective temperatures between
 log T$_{\rm{eff}}$ = 4.21 and log T$_{\rm{eff}}$ = 3.98 occurs in models where convective energy transport is negligible.

 The stability analysis of models with 200 M$_{\sun}$ and 250 M$_{\sun}$ using conventional boundary conditions 
 (Figs. \ref{200_modal_OBC} and  \ref{250_modal_OBC}) reveals results qualitatively similar to those for 150 M$_{\sun}$ 
 (Fig. \ref{150_modal_OBC}). With increasing mass (and luminosity) both the growth rate and the temperature range of the strange mode 
 instabilities increases to  4.4 > log T$_{\rm{eff}}$ > 3.93 (200 M$_{\sun}$)  and to  4.42 > log T$_{\rm{eff}}$ > 3.91 (250 M$_{\sun}$).
 On the other hand, both the growth rate and maximum effective temperature (log T$_{\rm{eff}}$ = 3.7) for the instability of the 
 convectively dominated models is almost independent of mass (and luminosity).

 Comparing the results of the stability analysis for boundary conditions consistent with the subsequent non-linear simulations 
 (Figs. \ref{150m_modal_nlb} - \ref{250m_modal_nlb}) with those based on the conventional outer boundary conditions
 (Figs. \ref{150_modal_OBC} - \ref{250_modal_OBC}), we recover counterparts of the strange mode instability and the instability for 
 log T$_{\rm{eff}}$ < 3.7. The latter is not affected by the boundary conditions, whereas the strange mode instability has become 
 stronger and affects more than a single mode. In addition, several dynamical monotonic instabilities are found. Obviously their 
 existence is due to the special choice of boundary conditions. The outer boundary conditions are in principle ambiguous, since the 
 outer boundary of the stellar model does not coincide with the boundary of the star. Therefore the physical relevance of instabilities
 caused by special boundary conditions remains an open question. We shall discuss this issue again in connection with the simulations of 
 the evolution of the instabilities into the non-linear regime and the final fate of unstable stellar models.

 Contrary to our study, the stability analysis of massive primordial stars by \citet{moriya_2015} has not revealed any instability 
 for log T$_{\rm{eff}}$ > 3.7, in particular not the strange mode instability extending up to log T$_{\rm{eff}}$ $\approx$ 4.5. 
 Whether the code used by these authors is not capable to identify strange mode instabilities or an operating error has prevented 
 the discovery of the instability needs further study. On the other hand, the instability for  log T$_{\rm{eff}}$ < 3.7 identified 
 by \citet{moriya_2015} is confirmed. However, as already noted, this instability occurs in convectively dominated models, 
 where the fundamental assumptions of the stability analysis are no longer valid. Thus the physical relevance of this instability
 is highly questionable and requires further investigations.

 Our study might also be regarded as a test for the strange mode instability mechanism proposed by \cite{glatzel_1994}. In the limit 
 of high luminosity to mass ratios and dominant radiation pressure, the dispersion relation for strange modes reduces to 
 \citep{glatzel_1994}:
 
 \begin{equation}
  \omega^{2} = \pm i 2  g \kappa _{\rm{\rho}} k
  \label{dis}
 \end{equation}
where $\omega$, $k$, $g$ and $\kappa _{\rm{\rho}}$ denote frequency, wave number, gravity and the logarithmic 
 derivative of the opacity with respect to density, respectively. For finite $k$ and $g$, this dispersion relation provides instability
 only if $\kappa _{\rm{\rho}}$ does not vanish. Thus apart from high L/M ratios and dominant radiation pressure, 
 $\kappa _{\rm{\rho}} \neq 0$ is an additional requirement for the occurrence of strange mode instabilities. For massive primordial 
 stars high L/M ratios and dominant radiation pressure prevail in all evolutionary phases. Close to the main sequence, 
 the effective temperatures are sufficiently high, such that the matter in the envelope is completely ionized and the opacity is 
 determined by electron scattering with $\kappa _{\rm{\rho}}$ = 0. As a consequence, strange mode instabilities should not 
 occur. However, as soon as in the post main sequence phase the effective temperatures becomes sufficiently low for helium to recombine, 
 bound-free transitions of helium determine the opacity implying  $\kappa _{\rm{\rho}} \neq 0$. Thus we expect strange mode instabilities
 to occur together with helium recombination below log T$_{\rm{eff}}$ $\approx$ 4.5. The findings of our study agree with these 
 predictions.

  \begin{figure}
    \centering $
 \Large
 \begin{array}{c}
   \scalebox{0.66}{ \input{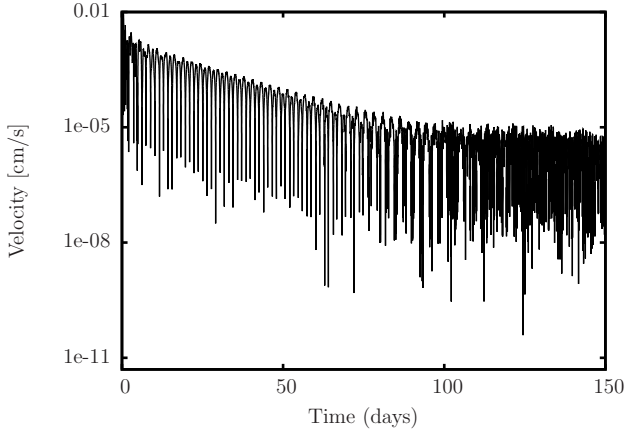} } \\
 \end{array}$
 
 \caption{Velocity of the outermost grid point as a function of time for a model with M = 
 150 M$_{\sun}$ and log T$_{\rm{eff}}$ = 4.6. Consistent with the linear analysis the simulation exhibits a 
 monotonic exponential increase saturating in the mildly non-linear regime. It is followed by an oscillatory 
 exponential decay and ends on the numerical noise level. 
 The final fate of this model is a new hydrostatic equilibrium with slightly different structure and parameters.}
 \normalsize
 \label{150m_40k}

 \end{figure}



  \begin{figure*}
    \centering $
 \LARGE
 \begin{array}{ccc}
   \scalebox{.455}{ \input{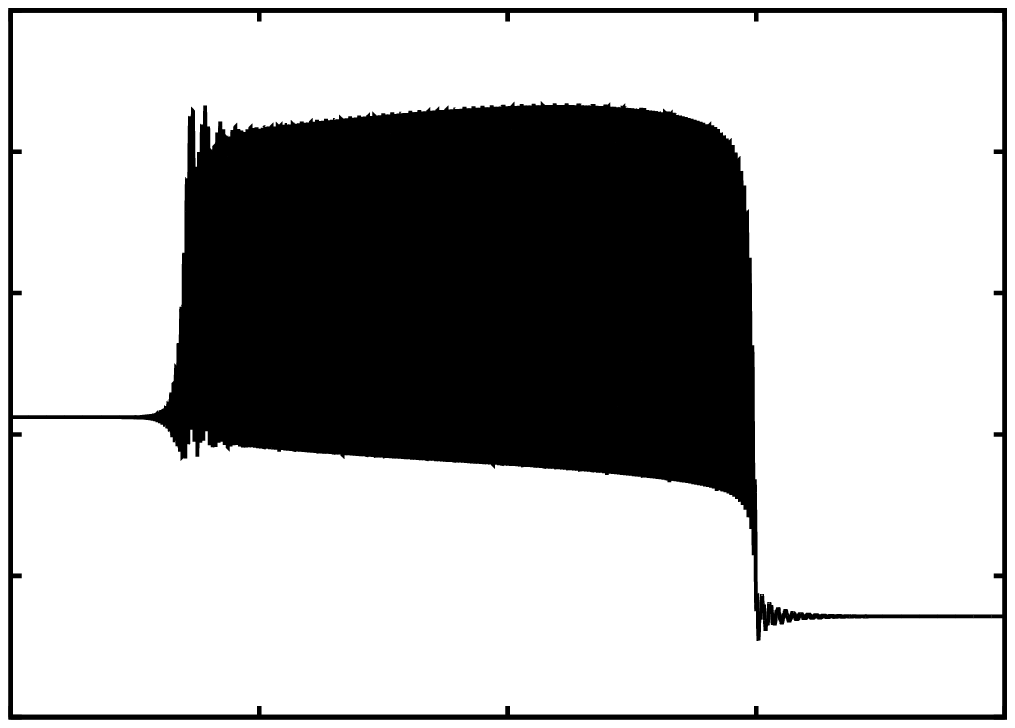} } 
   \scalebox{.455}{ \input{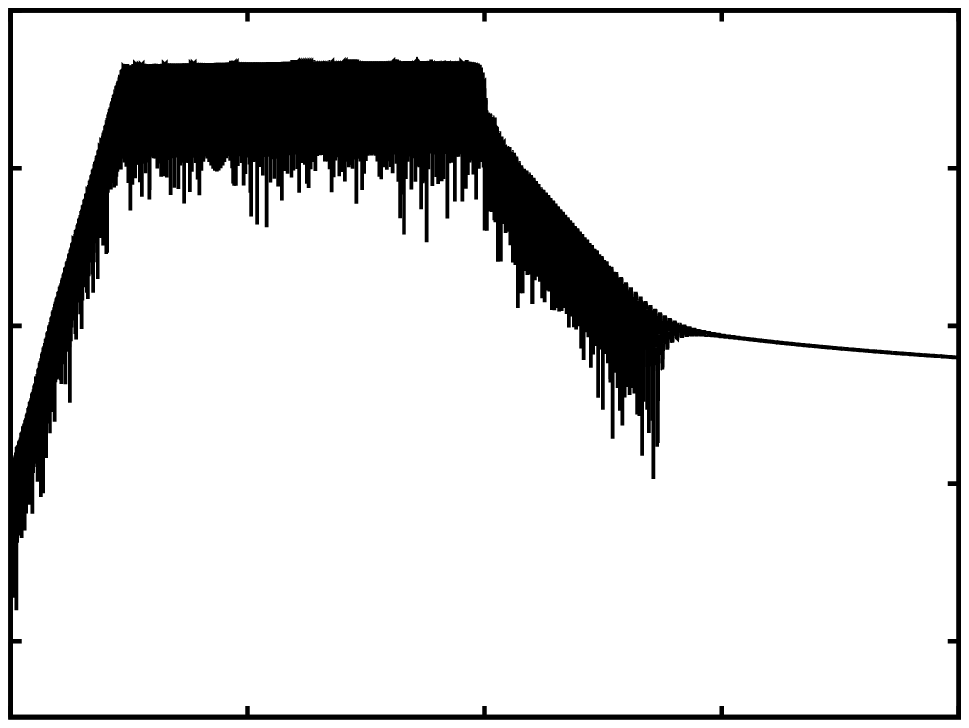} } 
   \scalebox{0.455}{ \input{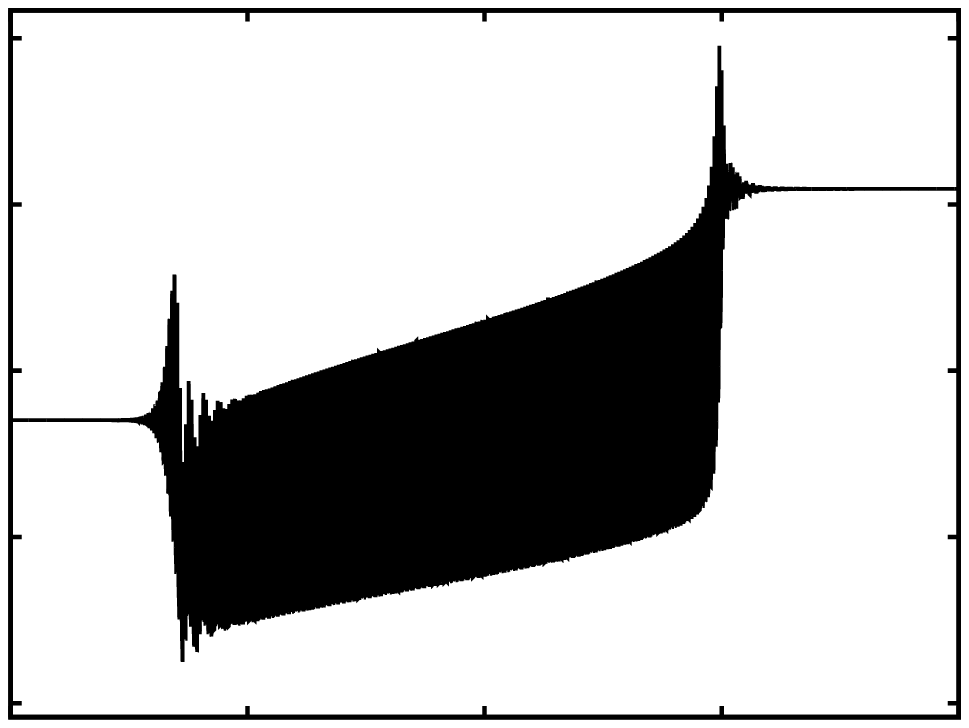} }                               \\
 
 \end{array}$
 
 \caption{ Evolution of instabilities into the non-linear regime for a model with M = 150 M$_{\sun}$ and log T$_{\rm{eff}}$ = 4.4.
 Radius (a), velocity (b) and temperature (c) at the outermost grid point are given as a function of time. The
 instability leads to a new hydrostatic state with modified structure and parameters.}
 \normalsize
 \label{150m_25k}

 \end{figure*}

   \begin{figure*}
    \centering $
 \LARGE
 \begin{array}{ccccccccc}
   \scalebox{.455}{ \input{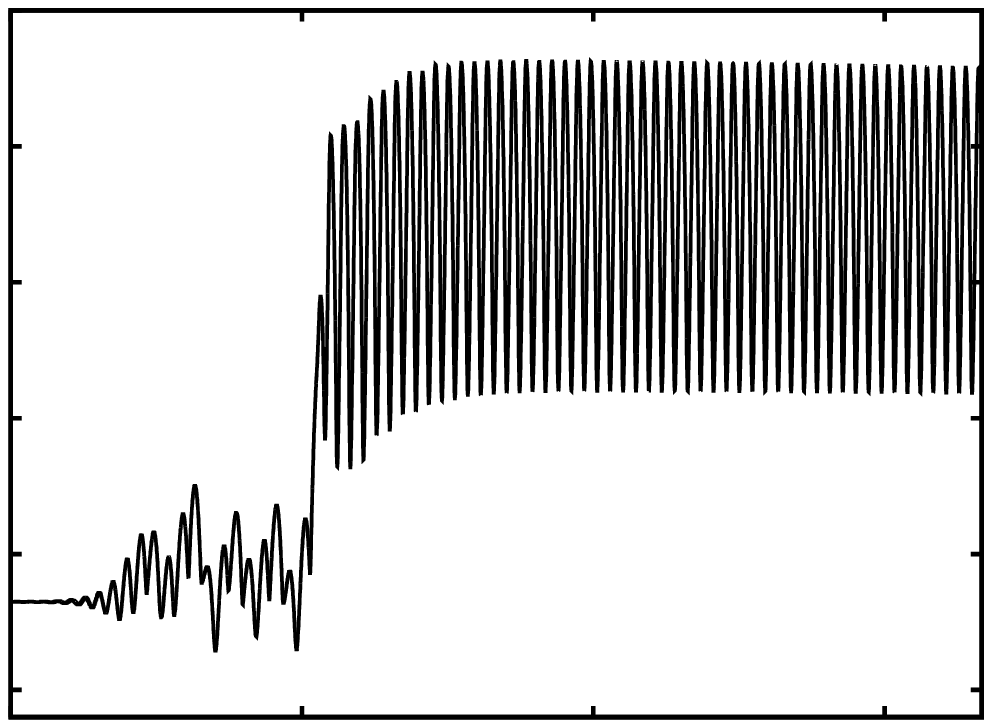} } 
   \scalebox{.455}{ \input{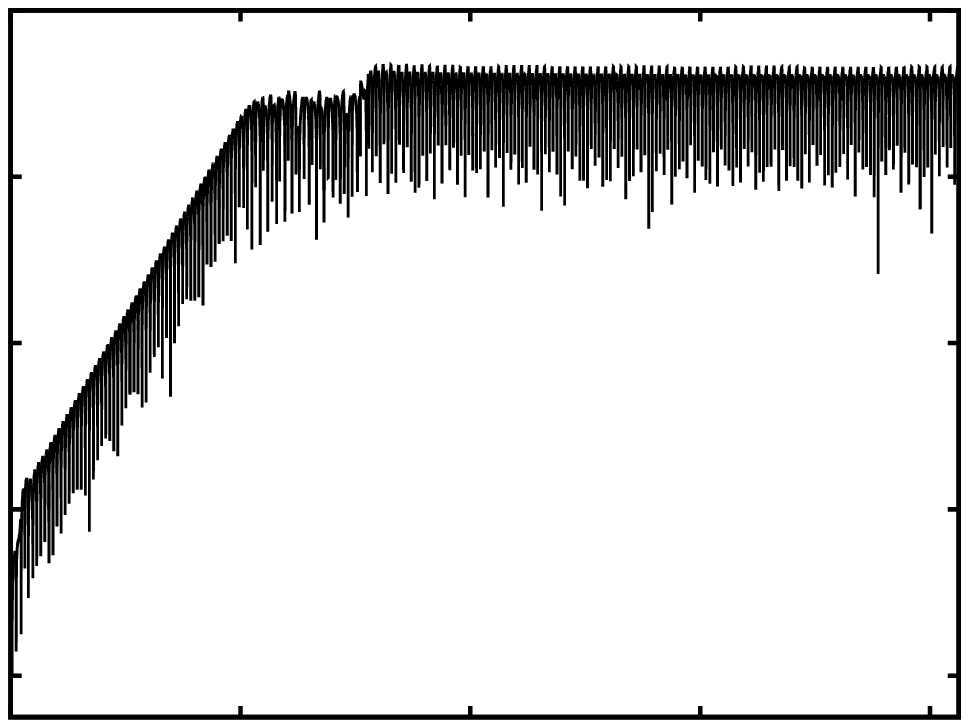} } 
   \scalebox{0.455}{ \input{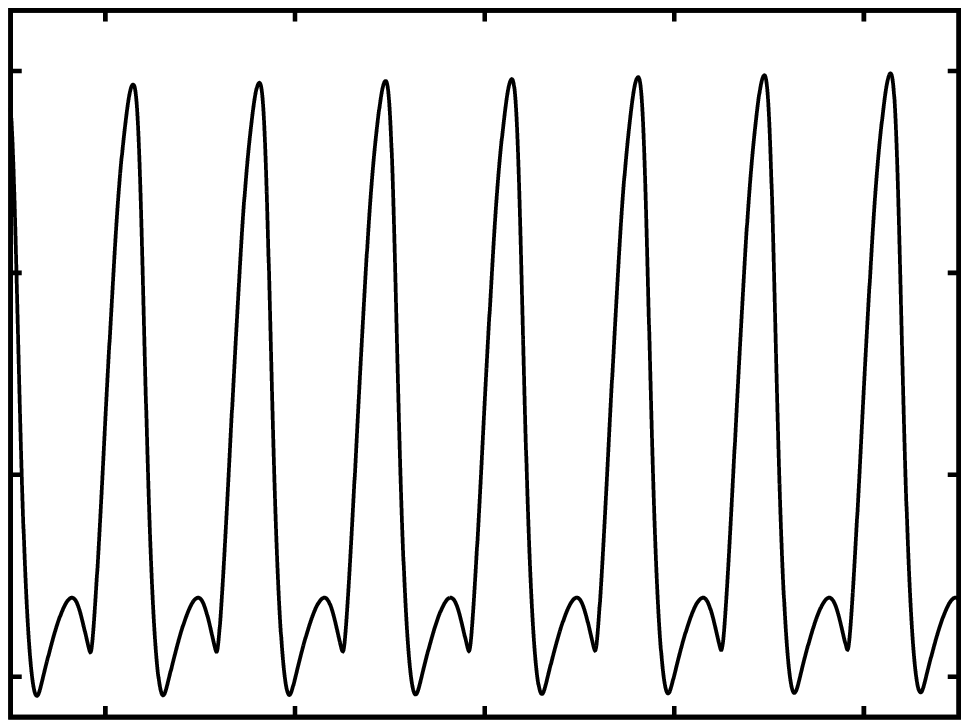} }                               \\
 \scalebox{.455}{ \input{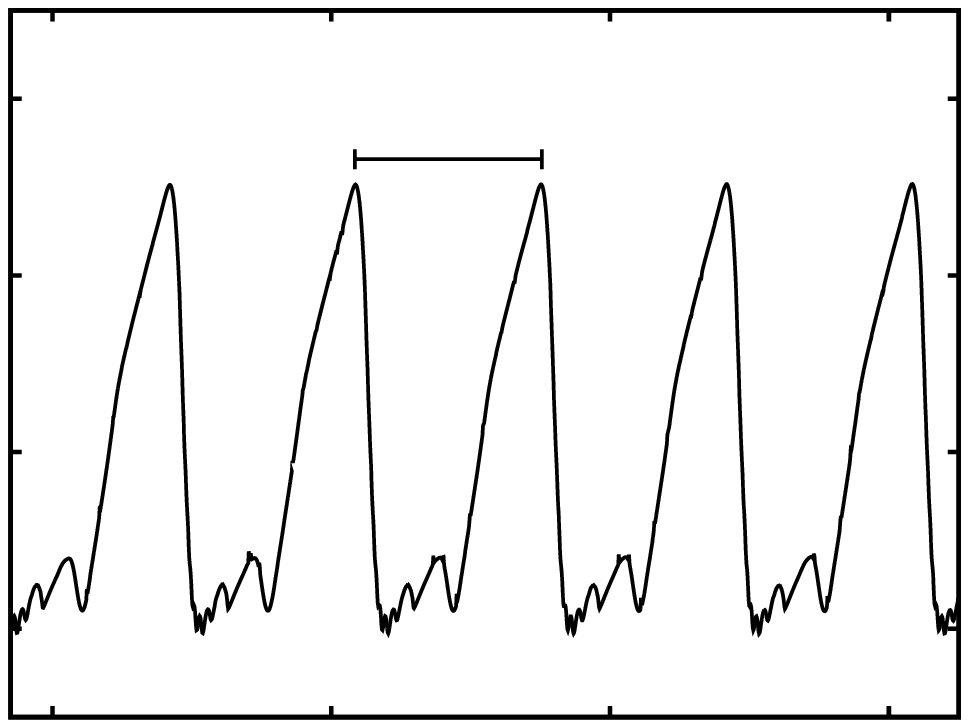} } 
   \scalebox{.455}{ \input{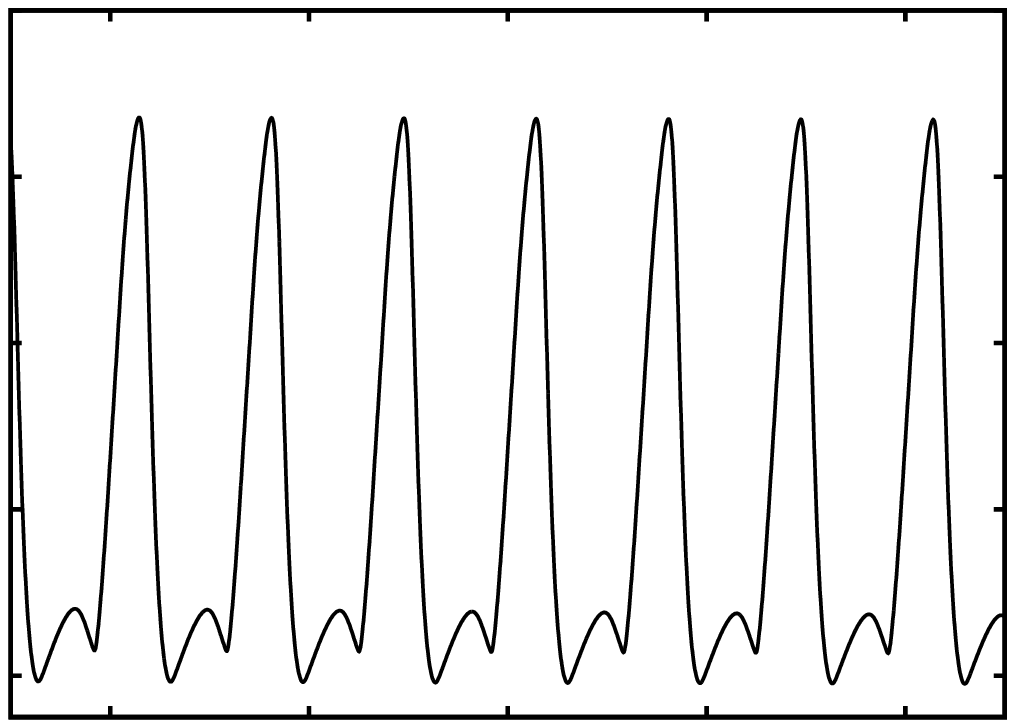} } 
   \scalebox{0.455}{ \input{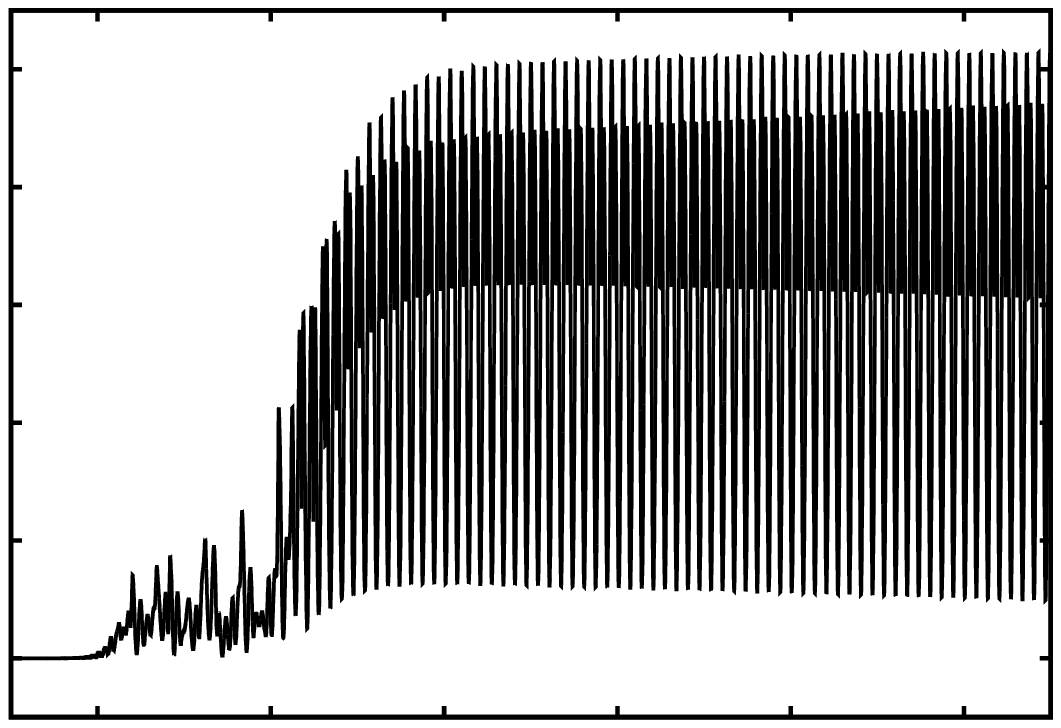} }                               \\
 \scalebox{.455}{ \input{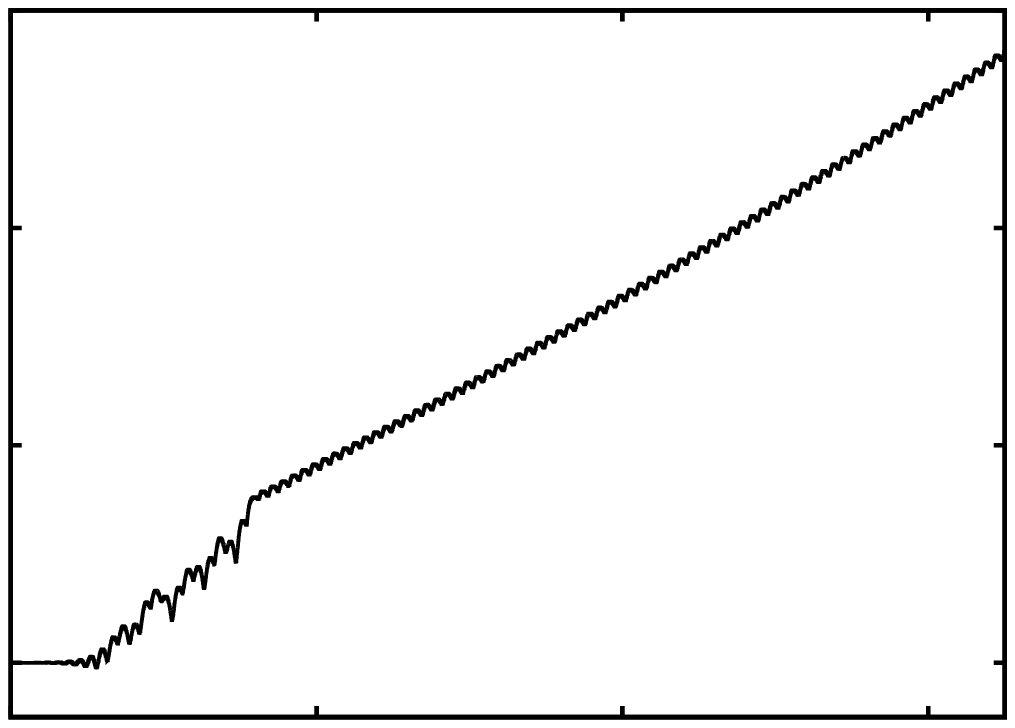} } 
   \scalebox{.455}{ \input{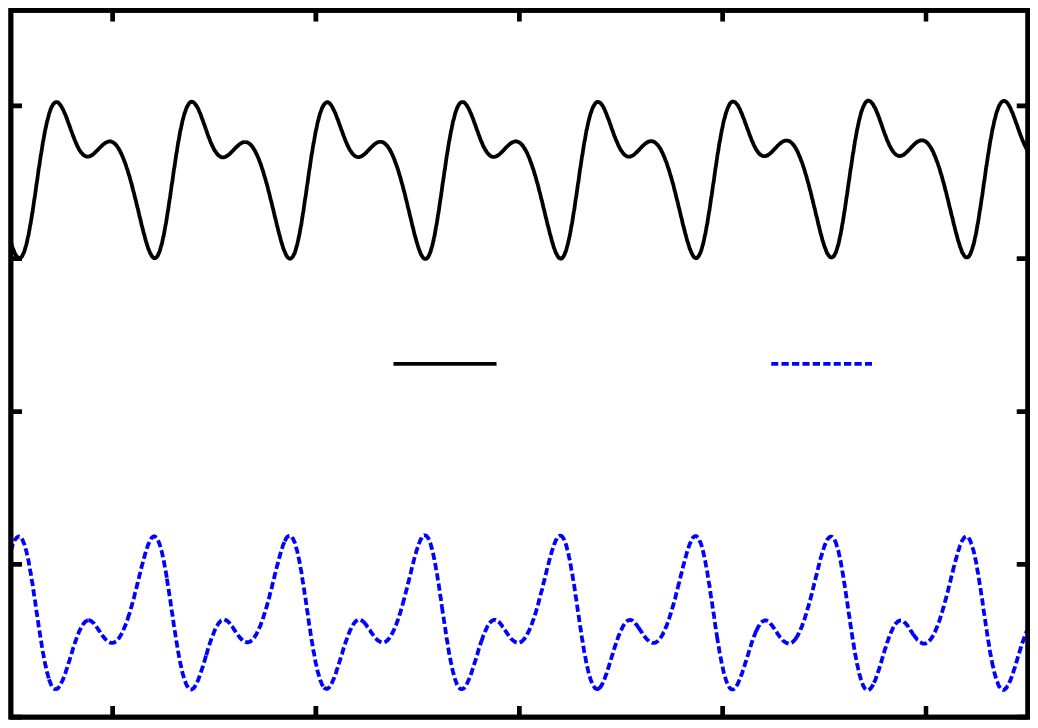} } 
   \scalebox{0.455}{ \input{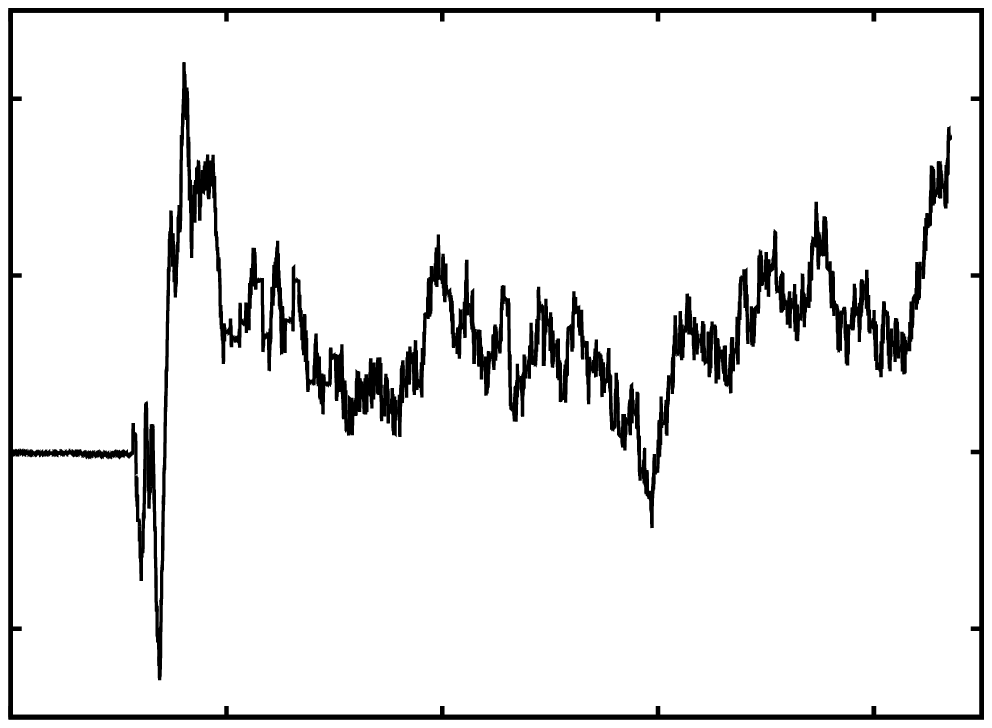} }                               \\
 
 \end{array}$
 
 \caption{Evolution of instabilities into the non-linear regime for a model having M = 150 M$_{\sun}$ and log T$_{\rm{eff}}$  = 4.2. 
 Radius(a), velocity (b), temperature (c) and density (e) at the outermost grid point, and the variation of the bolometric magnitude
 (d) are shown as a function of time. From the evolution of the velocity (b), 
 we deduce that the evolution of the instability starts from hydrostatic
 equilibrium with velocity perturbations of the order of 10 $^{-7}$ cm s$^{-1}$ superimposed,  undergoes the linear phase of 
 exponential growth and 
 saturates in the non-linear regime with an amplitude of 30 km$\,$s$^{-1}$. 
 Compared to the hydrostatic value, the mean radius is increased by $\approx$ 8 per cent in the non-linear regime.
 Some terms in the energy balance (with hydrostatic values subtracted) are given in (f)-(h) as a function of time. 
  Potential and internal energy (h) with almost identical modulus have opposite sign. They are bigger than the kinetic (f) 
  and the time integrated acoustic energy (g) by three and one orders of magnitude, respectively. The error in the energy balance
  is shown in (i). It is smaller than the smallest term in the energy balance by at least two orders of magnitude.}
 
 \normalsize
 \label{150m_15900K}

 \end{figure*}

   \begin{figure*}
    \centering $
 \LARGE
 \begin{array}{ccc}
   \scalebox{.455}{ \input{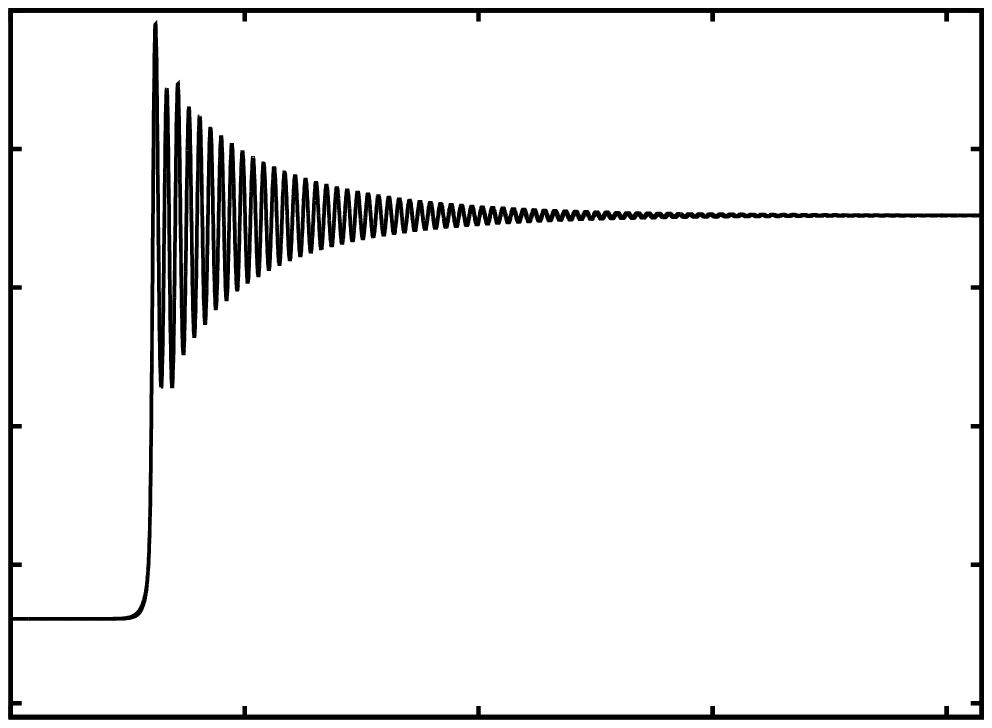} } 
   \scalebox{.455}{ \input{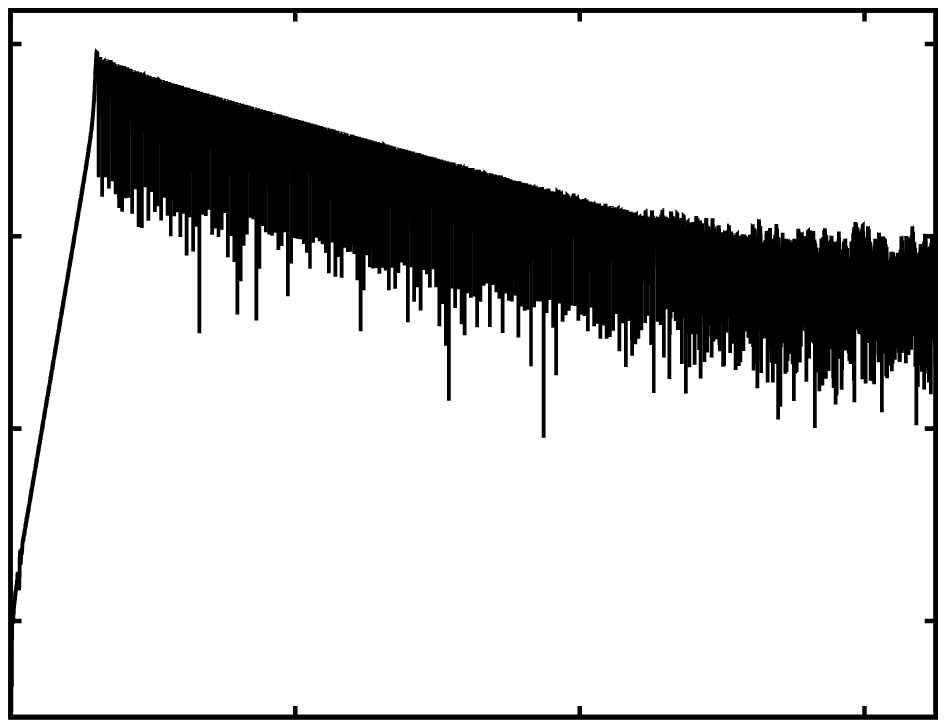} } 
   \scalebox{0.455}{ \input{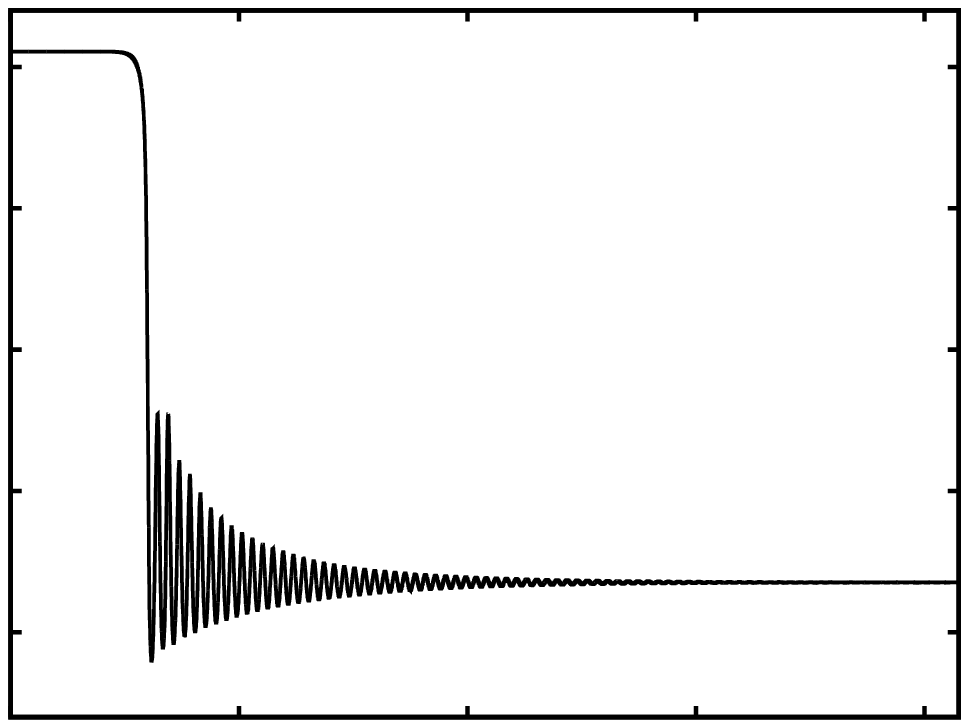} }                               \\
 
 \end{array}$
 
 \caption{ Same as Fig. \ref{150m_25k} but for a model having M = 150 M$_{\sun}$ and log T$_{\rm{eff}}$  = 4.0. 
 Note the presence of a monotonically unstable mode in the linear phase of exponential growth. The instability rearranges
 the stellar structure and a new hydrostatic state with 
 slightly increased radius and decreased effective temperature is attained.}
 \normalsize
 \label{150m_10k}

 \end{figure*}

   \begin{figure*}
    \centering $
 \LARGE
 \begin{array}{ccc}
   \scalebox{.455}{ \input{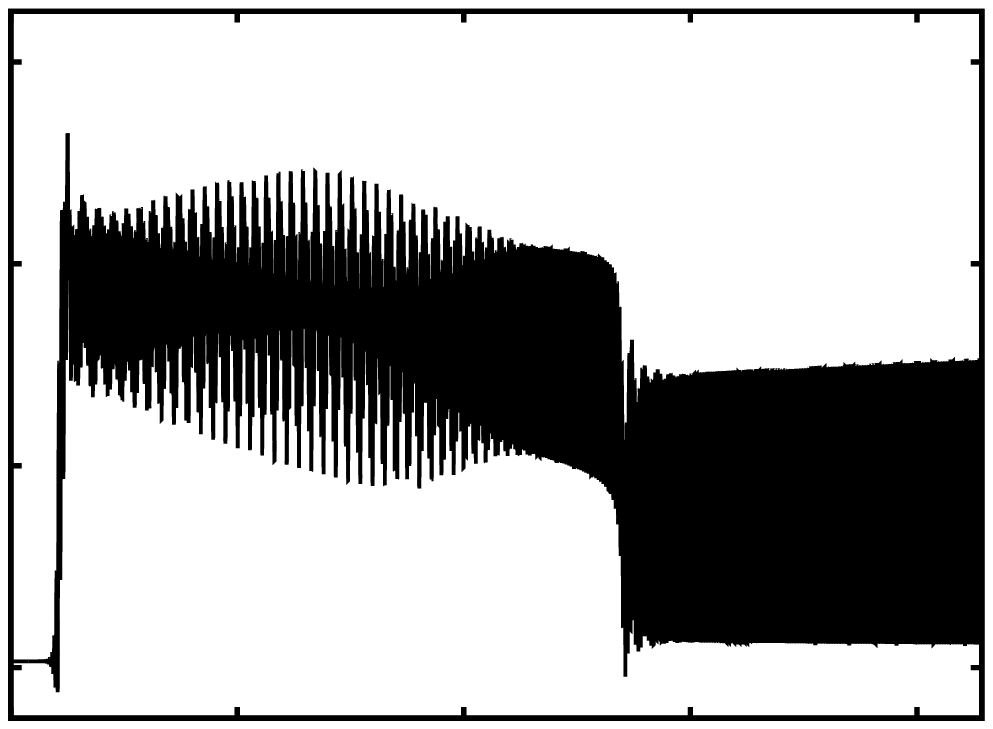} } 
   \scalebox{.455}{ \input{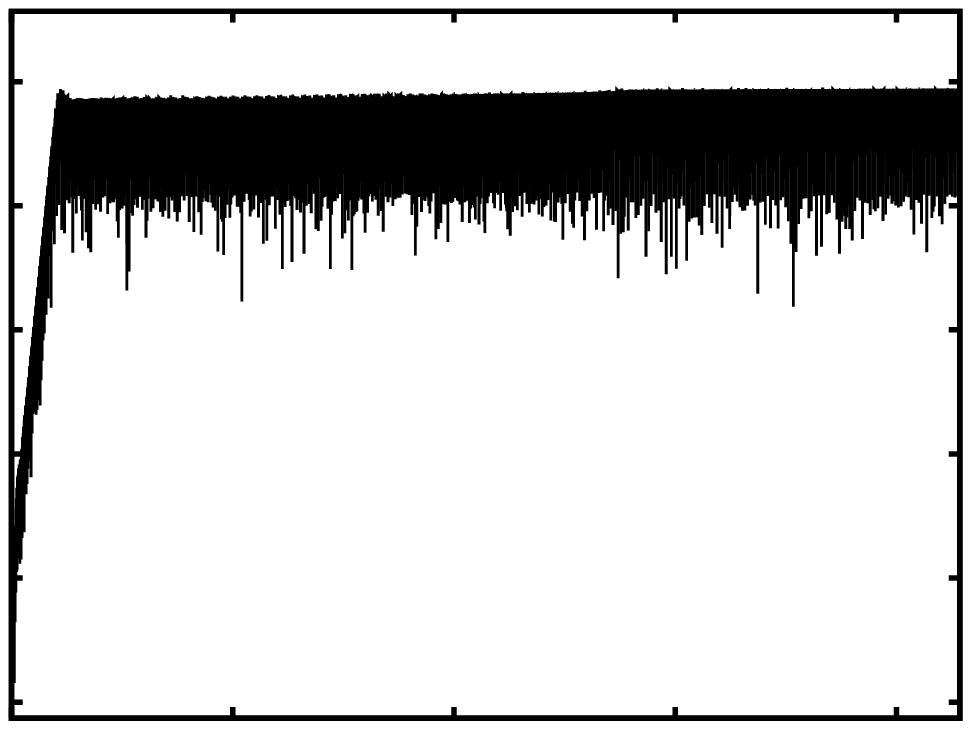} } 
  \scalebox{.455}{ \input{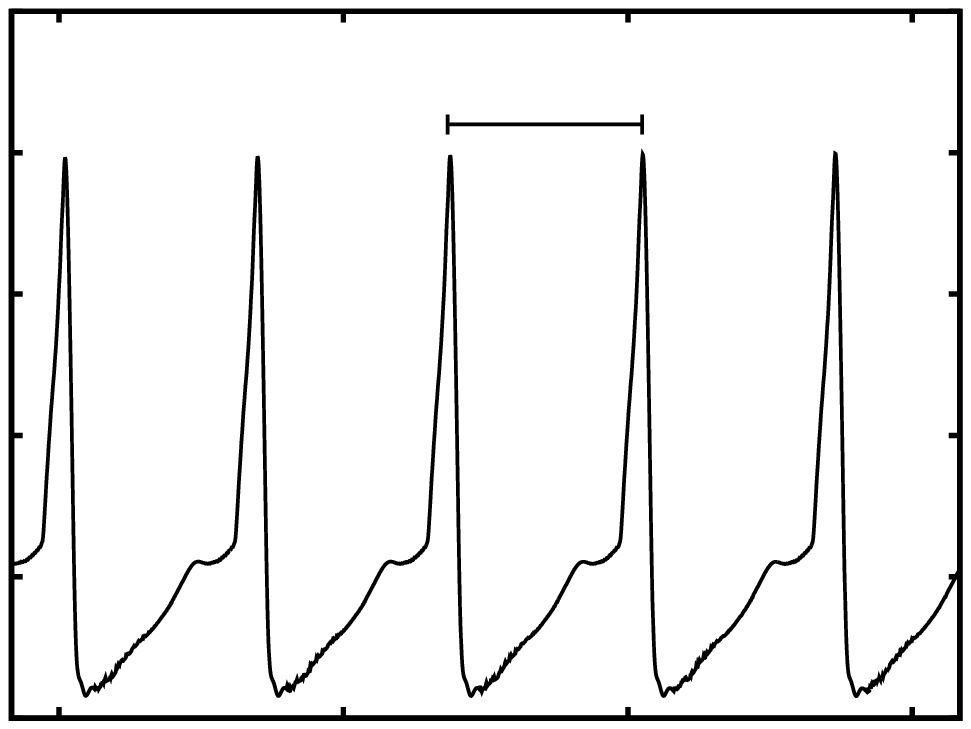} } \\
 \scalebox{.455}{ \input{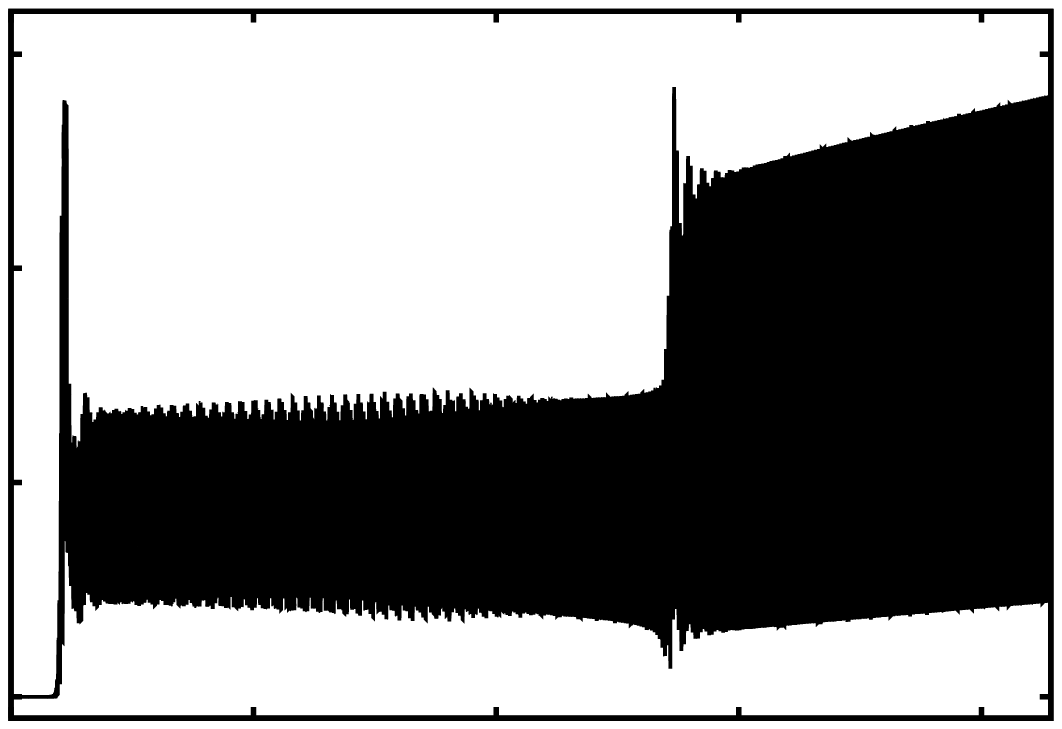} } 
 \scalebox{.455}{ \input{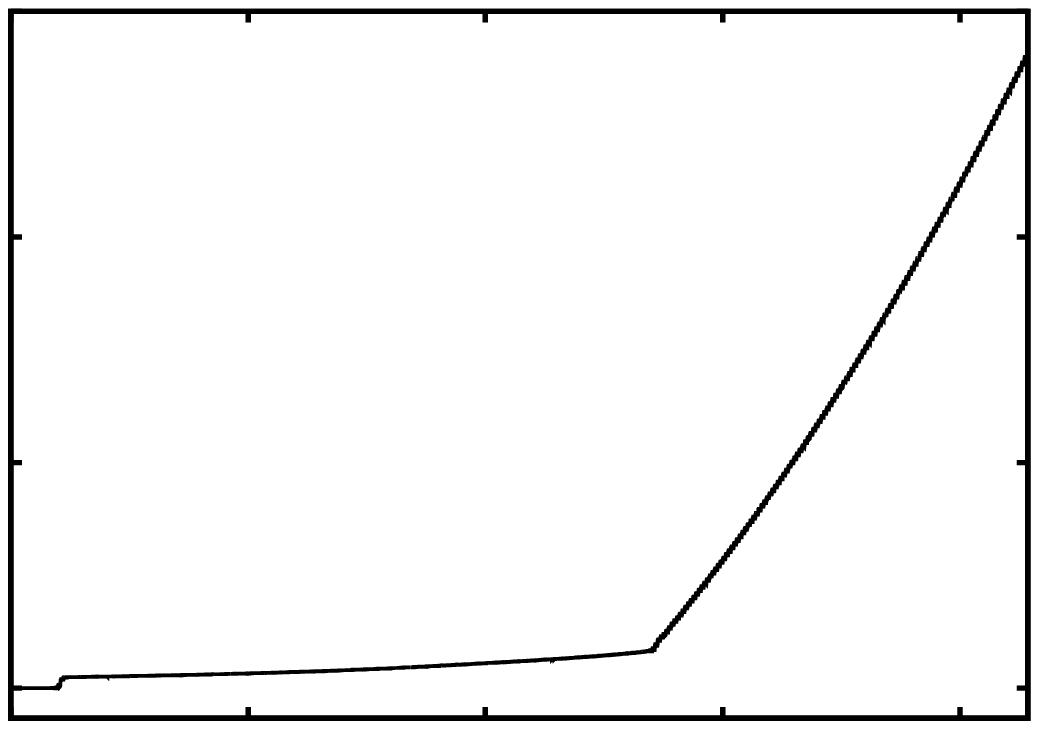} } 
   \scalebox{0.455}{ \input{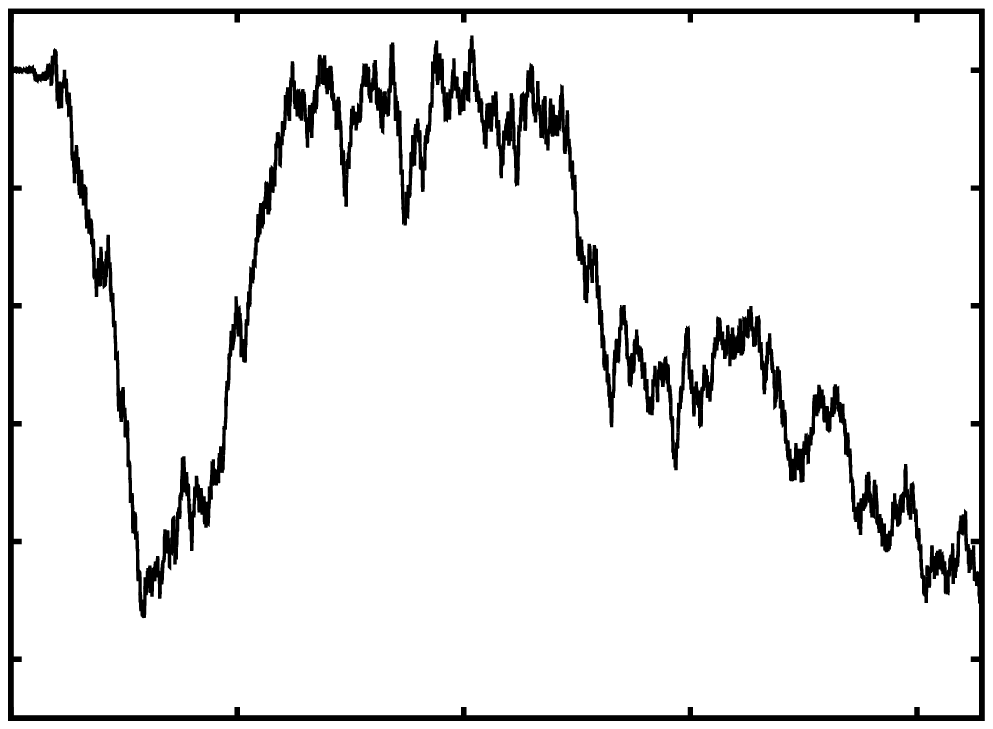} }                               \\
 \end{array}$
 
 \caption{Evolution of instabilities and finite amplitude pulsations for a model having M = 200 M$_{\sun}$ and log T$_{\rm{eff}}$  = 4.4. 
 As a function of time, the stellar radius, the velocity at the outermost grid point and the variation of the bolometric magnitude
 are displayed in (a)-(c), respectively. The velocity amplitude reaches 53 km$\,$s$^{-1}$ in the non-linear regime. 
 Kinetic energy, time integrated acoustic energy at the outer boundary and the error in the energy balance are 
 shown in (d)-(f), respectively.}
 \normalsize
 \label{200m_25k}

 \end{figure*}

   \begin{figure}
    \centering $
 \Large
 \begin{array}{c}
  \scalebox{0.65}{ \input{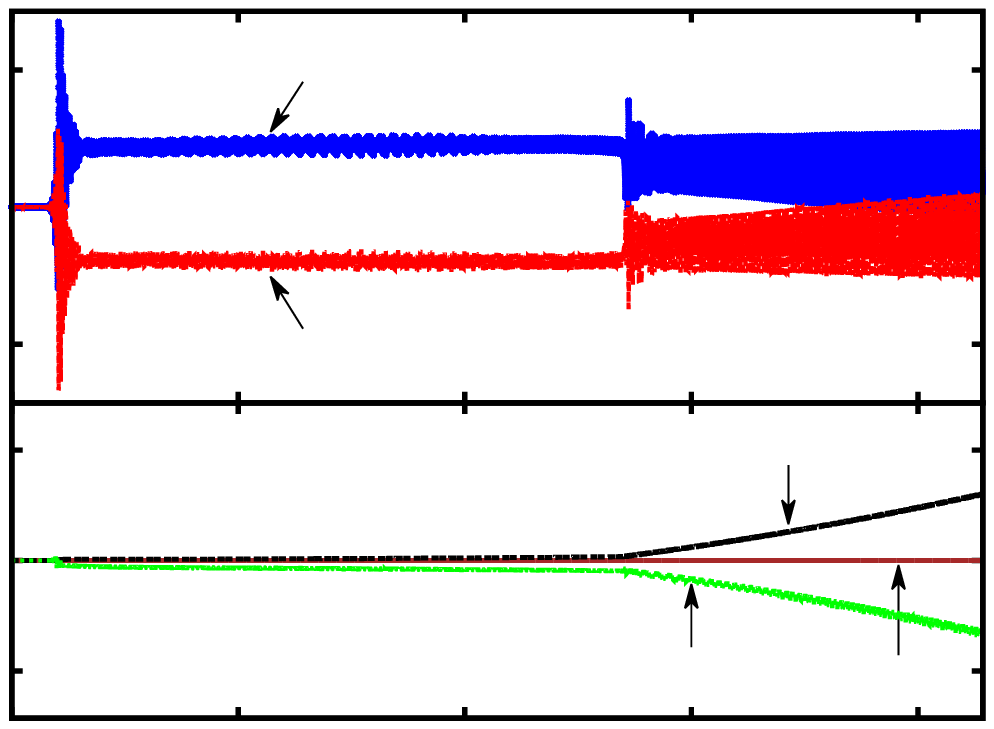} } \\
 \end{array}$
 
 \caption{ Potential, internal, kinetic, time integrated acoustic and time integrated thermal energy (at the outer boundary) 
 as a function of time for a model with M = 200 M$_{\sun}$ and log T$_{\rm{eff}}$ = 4.4. Potential and internal energies with almost 
 identical modulus and opposite sign are bigger than the kinetic energy by several orders of magnitude. Opposite signs and similar 
 moduli of the mean 
 slope of time integrated acoustic and thermal energies indicate that through the finite amplitude pulsations, thermal energy flux
 is transformed into acoustic energy flux.}
 \normalsize
 \label{200m_allen}

 \end{figure}

   \section{Nonlinear simulations}
  
  For selected unstable models, the evolution of the instabilities has been followed by numerical simulation into the non-linear regime 
  in order to determine their final fate. The numerical scheme used for this purpose is described in \citet{grott_2005}. It is fully 
  conservative with respect to energy, i.e., the energy balance is satisfied by the scheme intrinsically and locally. This property 
  is essential for the simulation of stellar instabilities and pulsations, since the kinetic energy and the time integrated 
  acoustic energy at the outer boundary are smaller than the dominant gravitational and internal energies 
  \cite[see equation 23 of][]{grott_2005} by several orders of magnitude for stellar pulsations. Thus for a meaningful determination 
  of the kinetic energy and time integrated acoustic energy (at the outer boundary) which we are interested in, an extremely high accuracy
  is required which can only be achieved by using a fully conservative scheme. This issue has meanwhile been discussed and emphasized 
  several times \cite[see, e.g.,][]{grott_2005, yadav_2016, yadav_2017, yadav_2017b}. As a consequence of conservativity, the numerical 
  scheme has to be implicit with respect to time. In the course of the non-linear evolution of instabilities, shock waves do occur. 
  They are represented by the introduction of artificial viscosity. For further details of the numerical treatment, we refer to 
  \citet{grott_2005}.

  One term occurring in the energy balance of the system corresponds to the time integrated acoustic energy at the outer boundary 
  \cite[see equation 23 of][]{grott_2005} representing the mechanical energy lost from the configuration by acoustic waves and 
  shocks. As discussed in a previous paper \citep{yadav_2017b}, there are phases of incoming and outgoing 
 acoustic fluxes during a pulsation cycle. As a consequence, the time integrated acoustic energy is a non-monotonic function.
 However, integrated 
 over one cycle the outgoing energy in general exceeds the incoming energy and on average the time integrated acoustic energy
 increases with time. Thus we obtain a mean slope of the integrated 
 acoustic energy, which corresponds to a mean mechanical luminosity of the system \cite[see][]{yadav_2017b}. 
 Assuming that this mean mechanical 
 luminosity is responsible for mass-loss of the star, we can estimate the mass-loss rate by comparing it to
 the wind kinetic
 luminosity $\frac{1}{2} \, \dot{M} \,v_{\infty} ^{2}$, 
 where $\dot{M}$ and $v_{\infty}$ are the mass-loss rate and terminal wind velocity, respectively 
 \citep[see also][]{grott_2005, yadav_2016, yadav_2017, yadav_2017b}. The terminal wind velocity is estimated by the escape velocity.
 In this way, an estimate is obtained in the following for the mass-loss rate from the mean slope of the time integrated acoustic energy.

   \subsection{Models with masses of 150 M$_{\sun}$}
 For models with masses of 150 M$_{\sun}$ (log L/L$_{\sun}$ = 6.6) four unstable configurations with effective temperatures of 
 log T$_{\rm{eff}}$ = 4.6, 4.4, 4.2 and 4.0 respectively have been selected for the numerical simulation of the evolution of 
 the instabilities into the non-linear regime. 
 
 At log T$_{\rm{eff}}$ = 4.6, a monotonic instability was identified using `minimum reflection boundary conditions' 
 (see Fig. \ref{150m_modal_nlb}), whereas this model turned out to be stable for the conventional boundary conditions 
 (see Fig. \ref{150_modal_OBC}). 
 Starting from numerical noise, the evolution of the instability (see Fig. \ref{150m_40k}) exhibits a linear phase of 
 monotonic exponential growth (with the growth rate predicted by the linear analysis) saturating at a weakly non-linear level 
 (below 0.01 cm/s in terms of the velocity amplitude). At this phase a slight modification of the structure of the model implies 
 stabilization. As a consequence, the model oscillates around its new equilibrium, i.e., we observe a damped oscillation ending - 
 in terms of the velocity amplitude - on the numerical noise level superimposed on the new hydrostatic equilibrium.

 At log T$_{\rm{eff}}$ = 4.4, an oscillatory instability was identified using `minimum reflection boundary conditions' 
 (see Fig. \ref{150m_modal_nlb}), whereas this model turned out to be stable for the conventional boundary conditions 
 (see Fig. \ref{150_modal_OBC}). 
 Starting from numerical noise, the evolution of the instability (see Fig. \ref{150m_25k})  exhibits a linear phase of 
 (oscillatory) exponential growth (with period and growth rate as predicted by the linear theory) saturating 
 after $\approx$ 150 days in the non-linear regime with a velocity amplitude of 38 km$\,$s$^{-1}$. After $\approx$ 600 days, the structure is 
 sufficiently modified to ensure stabilization and the configuration starts to oscillate around a new hydrostatic equilibrium with 
 an exponential decay of the superimposed perturbations. The decay of the velocity perturbations switches from oscillatory to 
 monotonic around $\approx$ 900 days.

 At log T$_{\rm{eff}}$ = 4.2, an oscillatory instability was identified independent of the boundary conditions 
 (see Figs. \ref{150m_modal_nlb} and \ref{150_modal_OBC}). After the linear phase of exponential growth non-linear 
 saturation is reached for this model after $\approx$ 600 days with a velocity amplitude of $\approx$ 30 km$\,$s$^{-1}$
 (see Fig. \ref{150m_15900K}b). Rather than a new hydrostatic equilibrium, finite amplitude pulsations are the consequence of 
 the instability for this model. An increase of the mean radius by $\approx$ 8 per cent in the non-linear regime is found 
 (see Fig. \ref{150m_15900K}a) implying the final non-linear pulsation period of 13.3 days to be higher than predicted by the 
 linear analysis. For illustration of the accuracy requirement and the numerical quality of the simulation, some terms 
 occurring in the energy balance \citep[see equation 23 of][]{grott_2005} together with its error are displayed in 
 Fig. \ref{150m_15900K}(f)-(i). Potential and internal energy (Fig. \ref{150m_15900K}h) have almost identical modulus and 
 opposite sign. They exceed the kinetic (Fig. \ref{150m_15900K}f) and the time integrated acoustic energy (Fig. \ref{150m_15900K}g)
 by three and one order of magnitude, respectively, whereas the error in the energy balance (Fig. \ref{150m_15900K}i) is smaller 
 than the smallest term in the energy balance by at least two orders of magnitude. From the mean slope of the time integrated 
 acoustic energy, we derive  7.7 $\times$ 10$^{-7}$  M$_{\sun}$ yr$^{-1}$ as an estimate for the mass-loss rate induced and
 driven by the pulsation.

 Similar to the model with log T$_{\rm{eff}}$ = 4.6, a monotonic instability was identified for log T$_{\rm{eff}}$ = 4.0
 using `minimum reflection boundary conditions' 
 (see Fig. \ref{150m_modal_nlb}), whereas stability or very weak instability was found for the conventional boundary 
 conditions (see Fig. \ref{150_modal_OBC}).
 Starting from numerical noise, the evolution of the instability (see Fig. \ref{150m_10k}) exhibits a linear phase of 
 monotonic exponential growth (with the growth rate predicted by the linear analysis) saturating in the non-linear regime 
 with a velocity amplitude below 10 km$\,$s$^{-1}$.  
 In this phase,  the structure becomes sufficiently modified to ensure stabilization and the configuration starts to oscillate
 around a new hydrostatic equilibrium with 
 an exponential decay of the superimposed perturbations. The decay of the velocity perturbations ends on the numerical noise level.

 From our study, we conclude that the model which is linearly unstable independent of the boundary conditions 
 (log T$_{\rm{eff}}$ = 4.2, Fig. \ref{150m_15900K}) finally exhibits finite amplitude pulsations (including mass-loss), whereas 
 instabilities caused by the boundary conditions only 
 (log T$_{\rm{eff}}$ = 4.6, 4.4, 4.0 ; Figs. \ref{150m_40k}, \ref{150m_25k}, \ref{150m_10k}) lead to modified hydrostatic equilibria. 
 Thus, with respect to the final fate of the model (hydrostatic equilibrium or finite amplitude pulsations with mass-loss) the 
 dependence on boundary conditions of the linear stability analysis becomes less important (see the discussion in section 4).

   \begin{figure*}
    \centering $
 \LARGE
 \begin{array}{ccc}
   \scalebox{.455}{ \input{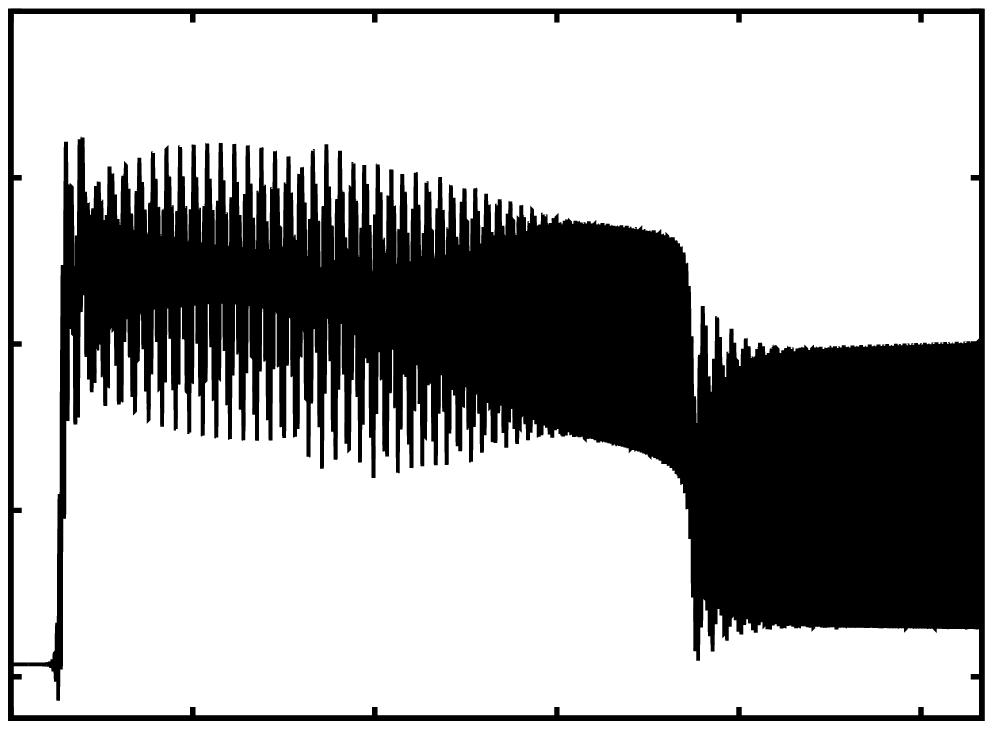} } 
   \scalebox{.455}{ \input{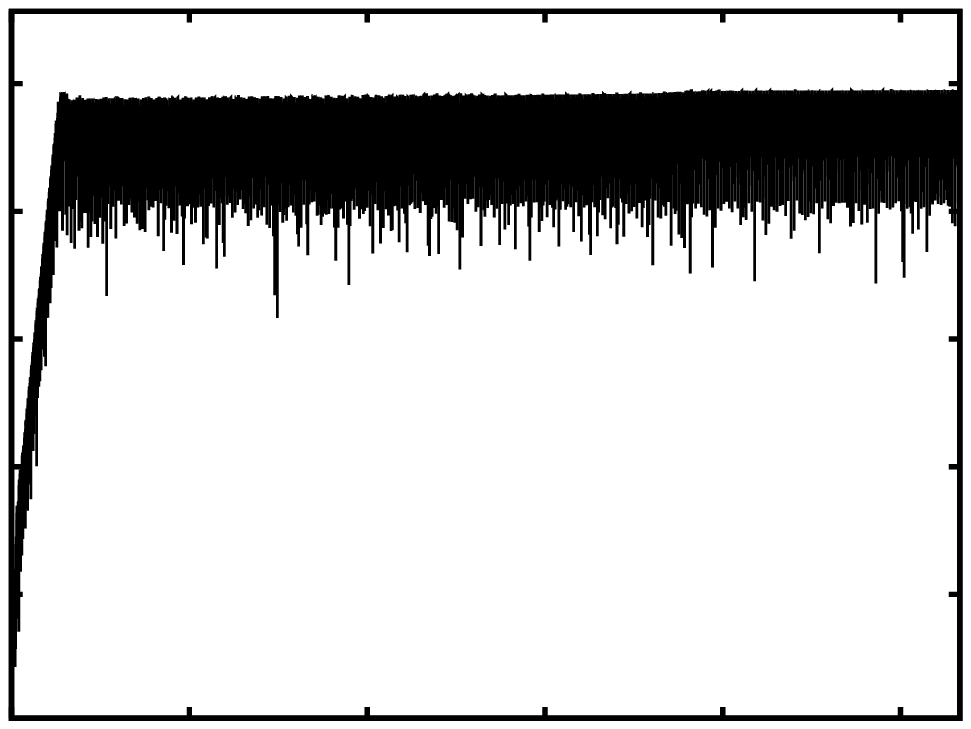} } 
   \scalebox{0.455}{ \input{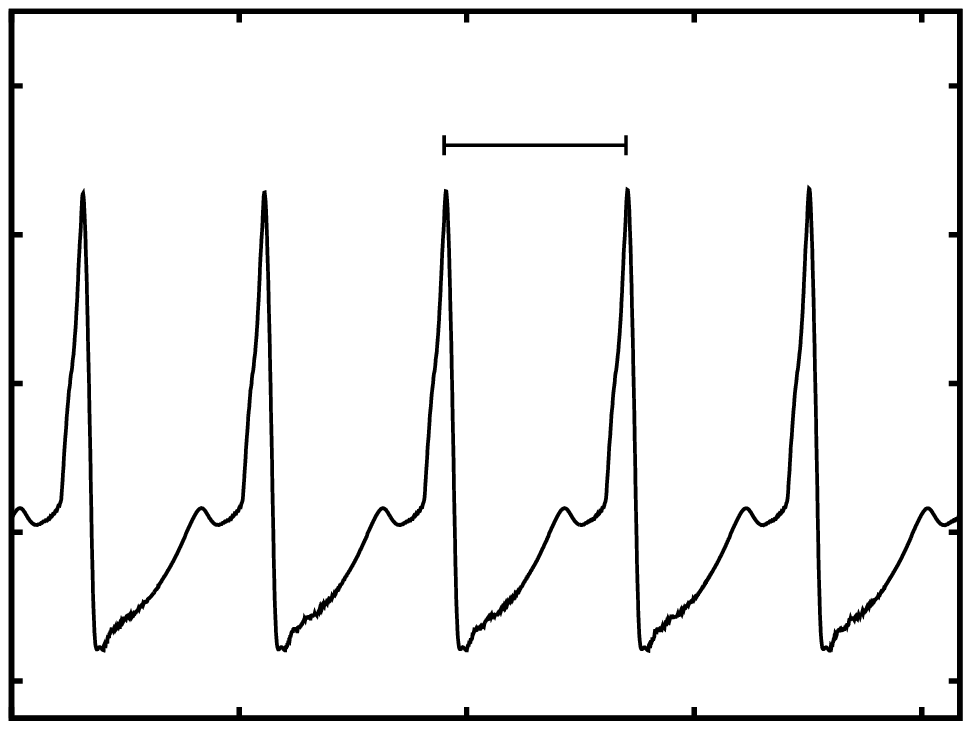} }                               \\
 \scalebox{.455}{ \input{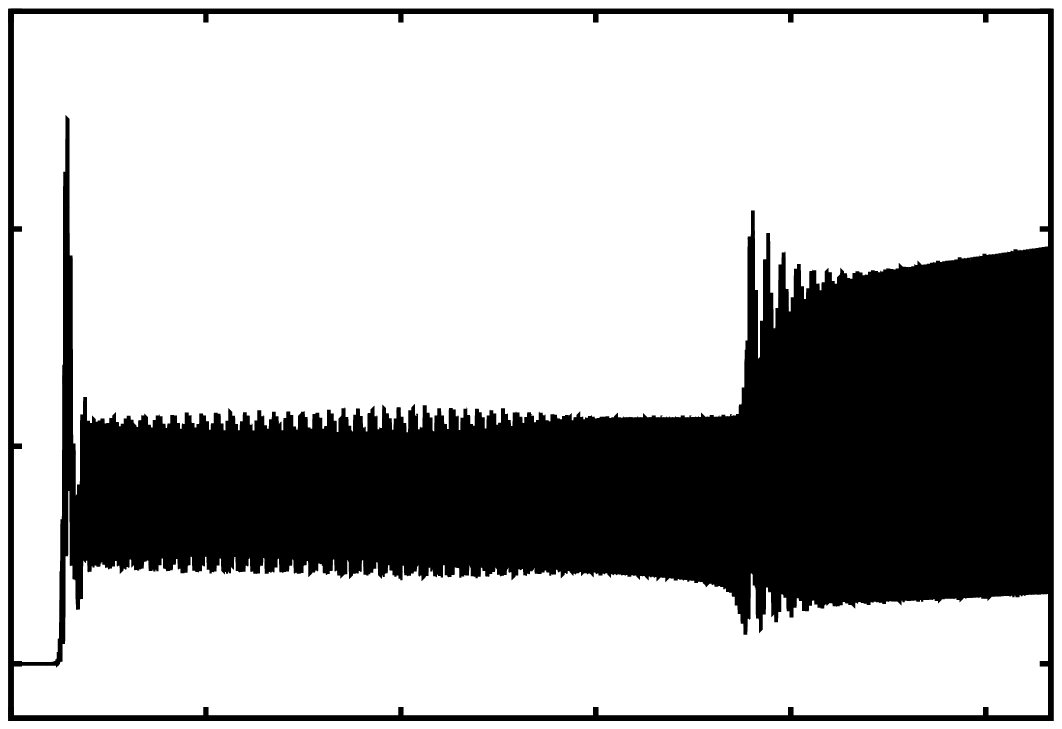} } 
   \scalebox{.455}{ \input{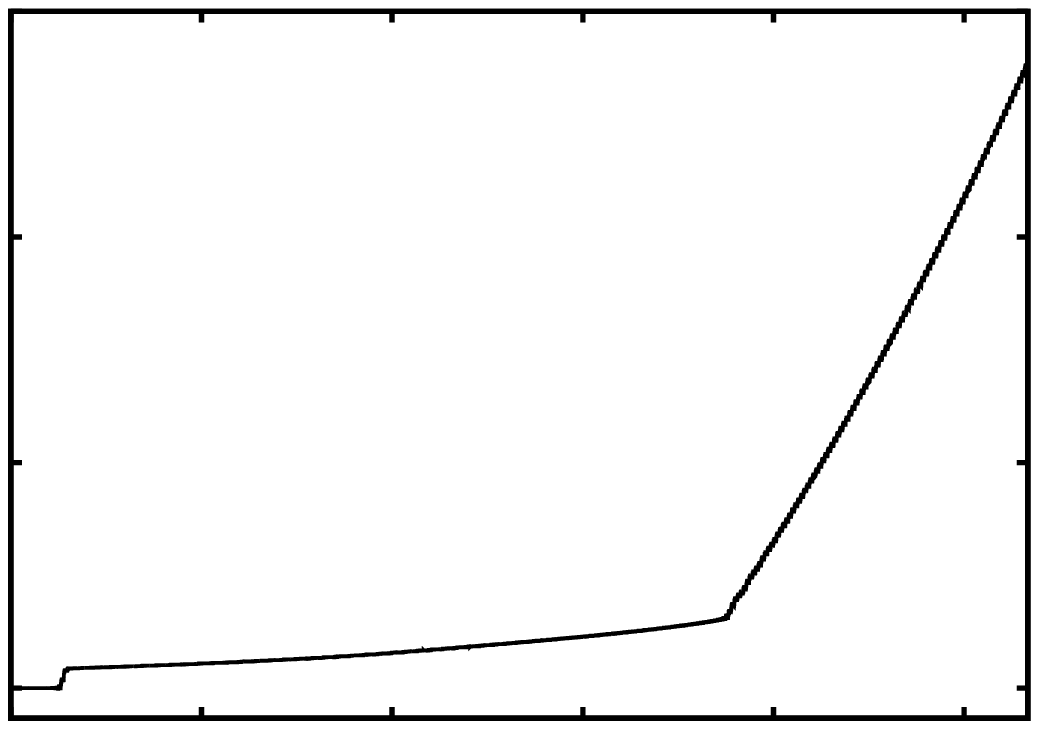} } 
   \scalebox{0.455}{ \input{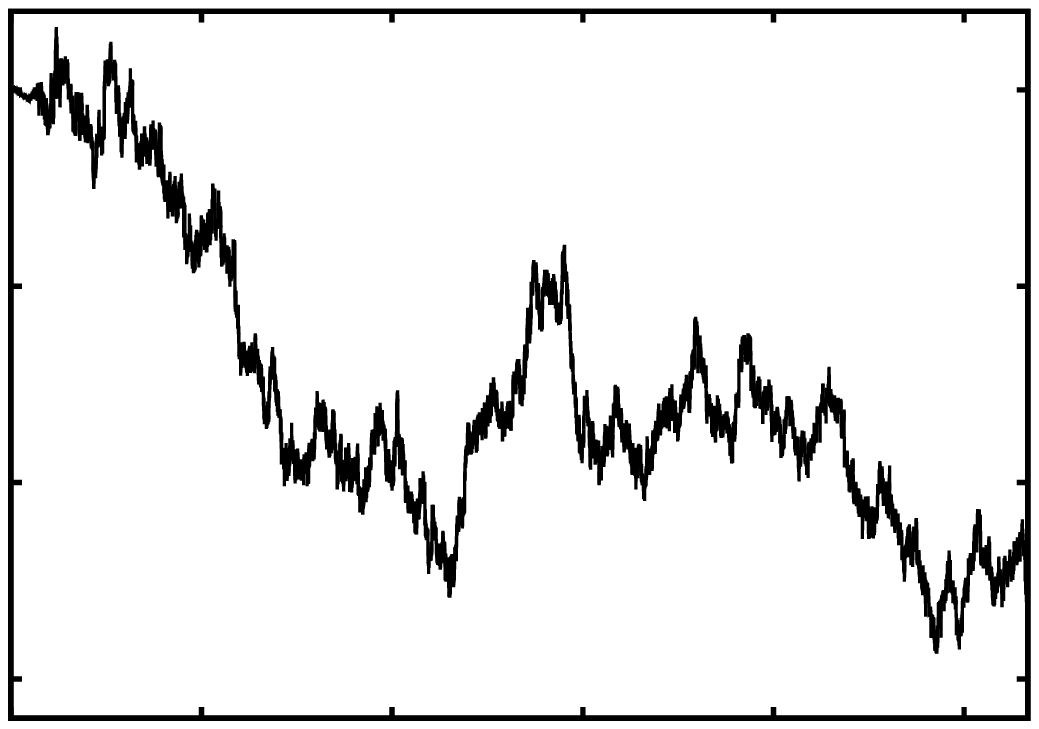} }                               \\
 
 \end{array}$
 
 \caption{Evolution of instabilities and finite amplitude pulsations for a model having M = 250 M$_{\sun}$ and log T$_{\rm{eff}}$  = 4.4. 
 As a function of time, the stellar radius, the velocity at the outermost grid point and the variation of the bolometric magnitude
 are displayed in (a)-(c), respectively. The velocity amplitude reaches 56 km$\,$s$^{-1}$ in the non-linear regime. 
 Kinetic energy, time integrated acoustic energy at the outer boundary and the error in the energy balance are 
 shown in (d)-(f), respectively.}
 \normalsize
 \label{250m_25k}

 \end{figure*}

  \subsection{Models with masses of 200 M$_{\sun}$ and 250 M$_{\sun}$}
  
  Similar to the unstable models with masses of 150 M$_{\sun}$ (log L/L$_{\sun}$ = 6.6), we have performed non-linear simulations
  of the evolution of instabilities 
  into the non-linear regime for unstable models with 200 M$_{\sun}$ and 250 M$_{\sun}$ (log L/L$_{\sun}$ = 6.77 and 6.88).  
  As examples, we present results for the models with log T$_{\rm{eff}}$  = 4.4 (which are unstable independent of boundary 
  conditions) in Figs. \ref{200m_25k} and \ref{250m_25k}.  
  For both 200 and 250 M$_{\sun}$, the instability leads to finite amplitude pulsations with a final period of 4.0 and 4.8 days and  
  velocity amplitudes of 53 and 56 km$\,$s$^{-1}$, respectively. From the mean slope of the time integrated acoustic energy, we estimate 
  a mass-loss rate of 5.4 $\times$ 10$^{-6}$ and  5.3 $\times$ 10$^{-6}$  M$_{\sun}$ yr$^{-1}$, respectively. We emphasize again 
  the importance of conservativity with respect to energy (see Figs. \ref{200m_25k}, \ref{250m_25k} d and f).

  A comparison of the various terms occurring in the energy balance \citep[see equation 23 of][]{grott_2005} provide inside into the 
  energy source for the pulsationally driven wind:  
  Potential, internal, kinetic, time integrated acoustic and time integrated thermal energy (at the outer boundary) 
 as a function of time are displayed in Fig. \ref{200m_allen} for the model with M = 200 M$_{\sun}$. 
 Potential and internal energies with almost 
 identical modulus and opposite sign are bigger than the kinetic energy by several orders of magnitude. Opposite signs and similar 
 moduli of the mean 
 slope of time integrated acoustic and thermal energies indicate that through the finite amplitude pulsations, thermal energy flux
 is transformed into acoustic energy flux.

 The results of our simulations are summarized in table \ref{table_periods}, where final pulsation periods and mass-loss rates 
 are given as a function of mass and effective temperature. Apart from a dependence on the strength of the underlying instability
 of the mass-loss rate, the latter seems to increase with mass and luminosity.

\begin{table*}
\centering
\caption{Pulsation periods and mass-loss rates for the primordial stellar models selected.}
\label{table_periods}
\begin{tabular}{|ccccl|}
\hline 
\begin{tabular}[c]{@{}c@{}} Mass\\    (M$_{\odot}$)\end{tabular} & \begin{tabular}[c]{@{}c@{}}log T$_{\rm{eff}}$\\     (K)\end{tabular} & \begin{tabular}[c]{@{}c@{}}Pulsation Period \\      (days)\end{tabular} & \begin{tabular}[c]{@{}c@{}}Mass Loss Rate\\     (M$_{\odot}$ yr$^{-1}$)\end{tabular} &  \\
\hline
\multirow{4}{4em}{  150} & 4.6 & Stable & --- \\ 
& 4.4 & Stable & --- \\
& 4.2 & 13.3 & 7.7 $\times$ 10$^{-7}$ \\
& 4.0 & Stable & --- \\
\hline
\multirow{2}{4em}{  200} & 4.4 & 4.0 &  5.4 $\times$ 10$^{-6}$ \\
& 4.2 & 20.0 & 2.58 $\times$ 10$^{-6}$ \\
\hline
\multirow{2}{4em}{250} & 4.4 & 4.8 & 5.3 $\times$ 10$^{-6}$ \\
& 4.2 & 25.3 & 3.5 $\times$ 10$^{-4}$ \\

\hline 
\end{tabular}
\end{table*}

  \section{Discussion and conclusions}

 A linear stability analysis has been performed for primordial post main sequence stellar models with masses between  
 150 and 250 M$_{\sun}$ (corresponding to luminosities between log L/L$_{\sun}$ = 6.6 and 6.88) covering the range of 
 effective temperatures between log T$_{\rm{eff}}$ = 4.80 and 3.62. The luminosity to mass ratios of these models lie 
 between 2.6 $\times$ 10$^{4}$ and 3.0 $\times$ 10$^{4}$ (solar units) and suggest the existence of strange mode
 instabilities with growth rates in the dynamical range which typically occur for luminosity to mass ratios
in excess of 10$^{3}$ \citep{gautschy_1990b, glatzel_1994}.

Contrary to previous investigations \citep{moriya_2015}, 
the expected strange mode instabilities have in fact been discovered, however only below an effective temperature of 
log T$_{\rm{eff}}$ $\approx$ 4.5. These findings are consistent with the predictions of a model for strange mode instabilities 
proposed by \citet{glatzel_1994}. According to it, in addition to high luminosity to mass ratios, a non-vanishing derivative 
of the opacity with respect to density is required for the existence of strange mode instabilities. For temperatures above 
  log T$_{\rm{eff}}$ $\approx$ 4.5, the matter in the stellar envelope is completely ionized and for primordial 
  chemical composition (Z = 0) only electron scattering contributes to the opacity. 
  Since the latter is constant, the derivative of opacity with respect to density vanishes and - according to the model 
  and in agreement with our findings - no strange mode instabilities do occur. On the other hand, if the effective 
  temperature falls below log T$_{\rm{eff}}$ $\approx$ 4.5, helium recombines and its bound-free transitions contribute to the 
  opacity resulting in a finite derivative of opacity with respect to density. Thus according to the model, strange mode 
  instabilities
  should appear together with helium recombination for temperatures below log T$_{\rm{eff}}$ $\approx$ 4.5, which agrees perfectly 
  with our results. We may thus take our results as a confirmation of the strange mode model introduced by \citet{glatzel_1994}.

  A second type of instabilities was identified for effective temperatures below log T$_{\rm{eff}}$ $\approx$ 3.7. 
  The energy transport
  in the envelopes of models affected by this instability is almost entirely due to convection. As any linear
  stability analysis developed
  so far does not contain a satisfactory treatment of convection, we refrain from further speculations concerning 
  possible implications 
  and consequences of this instability. In particular, contrary to \citet{moriya_2015} we have not performed 
  simulations of the evolution 
  of this instability into the non-linear regime.

  For selected stellar models, the evolution of strange mode instabilities into the non-linear regime was followed by numerical simulations. 
  Except for models, where the instability is caused by a special choice of boundary conditions (in these cases a modified hydrostatic
  equilibrium is the consequence of the instability) strange mode instabilities in primordial post main sequence models were found to lead to 
  finite amplitude pulsations with velocity amplitudes of the order of 50 km$\,$s$^{-1}$. Associated with these pulsations are acoustic energy fluxes
  capable of driving winds with mass-loss rates up to 3.5 $\times$ 10$^{-4}$ M$_{\odot}$ yr$^{-1}$. That these mass-loss rates are smaller 
  than those derived by \citet{moriya_2015} by at least two orders of magnitudes, is noteworthy. The post main sequence phase of the primordial stars studied
lasts for $10^4$ - $10^5$ years. Even the maximum mass-loss rates determined here would then influence the evolution of these
objects at most marginally.

  We emphasize that extremely high accuracy requirements have to be satisfied for the numerical treatment of stellar instabilities and 
  pulsations which can only be met by a with respect to energy fully conservative scheme. They are due to the fact, that the kinetic energies
  and the acoustic energy fluxes to be determined here, are smaller than the dominant gravitational and internal energies by several 
  orders of magnitude. In the simulations performed by \citet{moriya_2015} the kinetic energies reach a level (10$^{47}$ ergs) 
  which is typical for gravitational and internal energies but not for kinetic energies. We suspect that the numerical scheme 
  adopted by these authors is not conservative and the kinetic energies live on the numerical error of the gravitational and internal 
  energies. This would cast severe doubts on the reliability of this investigation, in particular on the high mass-loss rates claimed. 
  We note that the numerical calculations presented by \citet{appen_1970} suffer from the same problem.

Strange modes and associated instabilities are not restricted to radial perturbations and have been identified for 
 non-radial perturbations as well  \citep[see e.g.,][]{glatzel_1996, glatzel_2002}. Therefore we suspect the presence of 
 unstable strange modes in massive primordial stars  for non-radial perturbations too. 
A linear stability analysis for non-radial perturbations will be presented in a forthcoming study.

\section*{Acknowledgements}

APY gratefully acknowledges financial support through a SmartLink Erasmus Mundus Post-Doc fellowship.  




\bibliographystyle{mnras}
\bibliography{first} 







\bsp	
\label{lastpage}
\end{document}